\documentclass[oneside,
	preprint, 
	amsmath,
	amssymb,
	aps,
	prx,
	floatfix,
	11pt 
]{revtex4-2}
\usepackage[english]{babel}
\usepackage{graphicx}
\usepackage{xcolor}
\usepackage{bm}
\usepackage{physics}
\usepackage{booktabs}
\usepackage{subcaption}
\usepackage{titlesec} 
\usepackage{hyperref}
\usepackage{xr-hyper}
\usepackage{seqsplit}
\hypersetup{colorlinks=true,allcolors=blue}
\usepackage[separate-uncertainty=true]{siunitx}
\usepackage{setspace}
\usepackage[capitalise,nameinlink]{cleveref}
\crefname{video}{Video}{Videos}
\Crefname{video}{Video}{Videos}
\usepackage[right]{lineno}
\AtBeginDocument{\RenewCommandCopy\qty\SI}

\addto\captionsenglish{}

\makeatletter
\newcommand*{\addFileDependency}[1]{
    \typeout{(#1)}
    \@addtofilelist{#1}
    \IfFileExists{#1}{}{\typeout{No file #1.}}
}\makeatother

\DeclareCaptionLabelFormat{bf}{\textbf{(#2)}}
\newcommand\newsubref[1]{%
  \par\vspace{0.5em}
  \subref*{#1}%
}

\crefname{equation}{Equation}{Equations}
\Crefname{equation}{Equation}{Equations}
\creflabelformat{equation}{#2#1#3}

\usepackage{letltxmacro}
\usepackage{csquotes}

\makeatletter
\AtBeginDocument{%
  \LetLtxMacro\orig@nameref\nameref
  \DeclareRobustCommand*{\nameref}[1]{%
    \enquote{\textit{\orig@nameref{#1}}}%
  }%
}
\makeatother
\newcommand{\mytitle}{An adaptive fractional state links circuit mechanisms to cortical dynamics across the visual hierarchy}

\usepackage{algorithm}
\usepackage{algorithmic}
\makeatletter
\relax
\makeatother

\usepackage[parfill]{parskip}

\usepackage{setspace}
\usepackage{fontspec}
\usepackage{unicode-math}
\DeclareCaptionLabelFormat{continued}{#1 #2 continued}
\usepackage{setspace}
\begin{document}

\title{\mytitle}
\author{Brendan Harris}
\affiliation{School of Physics, The University of Sydney, Camperdown NSW 2006, Australia}

\author{Pulin Gong}
\thanks{Lead contact}
\email[Correspondence: ]{pulin.gong@sydney.edu.au}
\affiliation{School of Physics, The University of Sydney, Camperdown NSW 2006, Australia}
\affiliation{ARC Centre of Excellence for Integrative Brain Function, The University of Sydney, Camperdown NSW 2006, Australia}

\begin{abstract}
\vspace{0.5em}
Cortical circuits must respond flexibly to new inputs while integrating information about the past, yet the way in which neural activity reconciles these competing demands remains unclear.
Combining Neuropixels recordings from six mouse visual areas with mechanistic circuit modeling, we identify a dynamical regime in which heavy-tailed superdiffusive fluctuations coexist with long-range temporal dependence and oscillations.
We formalize this regime as the adaptive fractional (AF) state, using an effective mean-field theory in which spatial and temporal fractional derivatives capture heavy-tailed excursions and long-range memory, respectively.
We find that dynamical exponents characterizing the AF state vary systematically across the visual hierarchy: higher visual areas exhibit weaker superdiffusion and stronger temporal memory.
In the circuit model, this hierarchical shift emerges from a progressive weakening of effective inhibition.
These findings extend the classical hierarchy of timescales to a hierarchy of dynamical regimes and suggest that the AF state allows cortical areas to jointly express rapid responses and long-term temporal integration.

\end{abstract}

\date{\today}

\maketitle

\clearpage

\section{Introduction}
\label{sec:introduction}

Cortical circuits must respond rapidly to new inputs and transition flexibly between diverse activity patterns, while preserving information for long-term temporal integration.
These competing demands are reflected in neural dynamics at every level of cortical activity.
Individual neurons operate in a fluctuation-driven regime, with low firing rates~\cite{Barth2012}, sparse responses~\cite{Olshausen2004}, and mean membrane potentials well below the spiking threshold~\cite{Carandini2000,DeWeese2006}, driven by synaptic inputs that are large, correlated, and non-Gaussian~\cite{Okun2008,Graupner2013,Destexhe2001}.
Population activity mixes transient state changes and non-stationary oscillations with approximately scale-free $1/f$ fluctuations~\cite{Poil2012,DallaPorta2019}, and spike counts are super-Poissonian with variability growing as a power law with the length of an observation window~\cite{Churchland2010,Munn2020}.
In order to understand the neurophysiological mechanisms of cortical computation, a theory of the cortical working regime should explain how rapid fluctuations, persistent temporal dependencies, and oscillatory structure arise together in the same circuit.

Classical theories each account for a subset of these signatures.
In balanced networks, excitatory and inhibitory inputs largely cancel in the mean, leaving fluctuations that drive irregular, low-rate spiking~\cite{Vreeswijk1996,Brunel2000}.
Excitatory--inhibitory feedback can also produce population oscillations while individual neurons remain fluctuation-driven~\cite{Brunel2003,Brunel1999}, and spike-frequency adaptation introduces slow modulatory rhythms~\cite{Volo2019}.
To understand their collective dynamics, these circuit models are commonly reduced through mean-field approximations, in which the recurrent network is represented by an effective stochastic input to a single average neuron~\cite{Sompolinsky1988,BenYishai1995,Brunel2000,Wardak2021}.
Such reductions provide mechanistic explanations for firing rates, spike-count statistics, and collective oscillations~\cite{Brunel2003,Volo2018,Ahmadian2021}.
However, conventional mean-field descriptions approximate the collective input as Gaussian and Markovian~\cite{Brunel2000,Volo2019}, and so cannot capture dynamics beyond light-tailed fluctuations and finite correlation timescales.

Evidence for dynamics beyond the Gaussian--Markovian regime has accumulated across levels of neural organization.
Non-Gaussian statistics~\cite{Buzsaki2014} have been reported in bumpy synaptic fluctuations~\cite{DeWeese2006,Okun2008}, neuronal input distributions~\cite{Gu2019}, neural coding~\cite{Gutierrez2015}, synapse-level structural connectivity~\cite{Cossell2015}, and macroscopic brain dynamics~\cite{Roberts2015}.
Meanwhile, long-range temporal dependencies appear in microscopic spike timing~\cite{Bhattacharya2005} and macroscopic brain activity~\cite{LinkenkaerHansen2001,Palva2013}, expressed through approximately $1/f$ power spectra~\cite{He2014} and power-law scaling of spike-count variability~\cite{Churchland2010,Munn2020}.
Recent theoretical work has incorporated some of these properties separately: jump processes~\cite{Becker2024} and L\'evy-driven mean fields~\cite{Wardak2021} account for large synchronous inputs and heavy-tailed fluctuations, whereas colored-noise and memory-dependent formulations generate persistent temporal correlations~\cite{MorenoBote2004,Fourcaud2002}.
Yet no existing framework allows non-Gaussian fluctuations, long-range memory, and oscillations to coexist while remaining separately tunable and experimentally measurable.

This fundamental gap limits our understanding of how neural dynamics are organized across cortex.
Along sensory hierarchies, higher areas integrate information over progressively longer intrinsic timescales~\cite{Murray2014,Rudelt2024,Shi2025} and differ systematically in their excitatory and inhibitory composition~\cite{Kim2017,Wang2020,FroudistWalsh2023}.
However, these hierarchical differences in dynamics have been summarized mainly by a single characteristic timescale, which specifies how long activity persists but not how that persistence relates to the magnitude and scaling of rapid fluctuations.
It remains unclear how temporal persistence and rapid fluctuations vary together across cortical areas, and which circuit interactions establish the working regime of each area.

Here, we combine Neuropixels recordings across six areas of the mouse visual cortex with a biophysical circuit model and a generalized mean-field theory.
We identify a cortical working regime that we term the `adaptive fractional' (AF) state; in this state, heavy-tailed superdiffusive excursions, subdiffusive long-range temporal memory, and oscillations coexist.
These components can be compactly described in terms of fractional-order dynamics---generalizations of integer-order equations whose non-integer derivatives capture nonlocal jumps and power-law memory~\cite{Metzler2000}---together with neural adaptation, which motivates the term `adaptive fractional'.
We formalize this description by generalizing the fractional neural sampling model~\cite{Qi2022a}, adding simultaneous fractional spatial and temporal derivatives.
The spatial fractional order controls the prevalence of large non-Gaussian excursions, the temporal fractional order controls long-range memory, and an auxiliary momentum term generates oscillations.
In appropriate limits, our bi-fractional model recovers established Gaussian~\cite{Brunel2000}, L\'evy-driven~\cite{Wardak2021}, and oscillatory~\cite{Aitchison2016} descriptions of collective neural dynamics, together with the fractional Langevin description of power-law memory~\cite{Li2017b}, within a common framework.
Importantly, our bi-fractional model dissociates the statistics of rapid, state-changing excursions from the persistence of activity over longer timescales while preserving oscillatory structure, and it reproduces the subthreshold and spiking signatures observed in both the circuit model and the recordings.
The AF state thus links dynamical signatures, which can be experimentally measured across multiple scales of cortical activity, to the space-fractional and time-fractional operators of a compact diffusive theory.

Crucially, the dynamical exponents defining the AF state vary systematically along the visual hierarchy.
Higher visual areas exhibit stronger long-range temporal memory and weaker superdiffusion, producing a coordinated shift in dynamical state that is most pronounced in superficial layers.
Circuit modeling links this hierarchical shift to systematic changes in the effective balance between excitation and inhibition, providing a mechanistic interpretation consistent with known anatomical gradients~\cite{Kim2017,D’Souza2016}.
The AF state thus offers both a unifying view of the cortical working regime and a new perspective on how cortical circuits balance rapid exploration against sustained temporal integration across the hierarchy.

\section{Results}

\subsection{Cortical activity combines superdiffusion, long-range memory, and oscillations}

To determine whether non-Gaussian fluctuations, long-range memory, and oscillations coexist in cortical activity, we analyze publicly available Neuropixels recordings from the awake mouse visual cortex~\cite{Bennett2025} (illustrated in \cref{subfig:schematic_experiment}).
We restrict the analysis to spontaneous activity, recorded during presentations of a uniform gray screen, to minimize stimulus-locked fluctuations.
After applying the quality-control criteria detailed in \nameref{sec:methods}, the dataset comprises $69$ sessions across $41$ mice, with local field potential (LFP) and spiking activity sampled across cortical depth in six visual areas.
We focus initially on layer 2/3 of the primary visual cortex (\texttt{VISp}), which is characterized by strong recurrent connectivity~\cite{Hage2022} and spike-frequency adaptation in pyramidal neuron populations~\cite{Fernandez2018}; representative LFP and population spiking activity are shown in \cref{subfig:brain_trace}.

LFP fluctuations are superdiffusive at short time lags.
We compute the mean absolute deviation (MAD) of the increments of an LFP time series $x(t)$ over a time lag $\tau$, $\text{MAD}(\tau) = \langle | x(t + \tau) - x(t) | \rangle_t\,,$ where $\langle \cdot \rangle_t$ denotes the average over time.
The scaling relation MAD~$\propto \tau^a$ defines the `diffusion exponent' $a$, which takes the value $a = 0.5$ for Brownian motion at small lags, so that $a > 0.5$ indicates superdiffusion.
We use the MAD rather than the mean squared displacement or detrended fluctuation analysis~\cite{Kantelhardt2001} because averaging absolute rather than squared increments reduces sensitivity to extreme values and remains well defined for heavy-tailed increments that have a finite first moment but divergent variance (\nameref{sec:methods}).
The LFP MAD follows an approximate power law at short lags before flattening at longer lags (\cref{subfig:empirical_mad}), reflecting confinement that prevents the LFP from wandering indefinitely.
Fitting over the short-lag scaling regime ($\SI{0.8}{\milli\second} \le \tau \le \SI{8}{\milli\second}$) gives a median diffusion exponent of $a = 0.62$ (95\% CI: $[0.60, 0.64]$ across sessions), so the rapid fluctuations of the LFP depart from the diffusive scaling expected of a Gaussian--Markovian process.

What statistical properties generate these superdiffusive dynamics?
For a Gaussian process, superdiffusion requires positively correlated increments and steep power spectra: writing the power spectral density as PSD~$\propto f^b$ and terming $b$ the `spectral exponent', persistent superdiffusive motion requires $b \le -2$~\cite{Krapf2019}, with unconfined Brownian motion having $b = -2$.
However, we find that LFP fluctuations have shallow exponents, with $b > -2$.
Since LFP spectra mix oscillatory peaks with broadband components, we decompose the PSD into oscillatory and aperiodic parts (\cref{subfig:empirical_psd}), revealing $\theta$ and $\gamma$ oscillations at \SI{7}{\hertz} and \SI{55}{\hertz} superimposed on an aperiodic component with $b = -1.75$ (95\% CI: $[-1.79, -1.72]$).
This shallow exponent implies the presence of anticorrelated increments that, in a Gaussian process, would produce subdiffusion rather than the superdiffusion observed in the MAD.

This apparent contradiction can be resolved by non-Gaussian statistics, as heavy-tailed increments generate superdiffusion without positive memory.
To test whether the LFP fluctuations are heavy-tailed, we pool the step-size distribution across layer 2/3 channels and compare it against phase-randomized Fourier-transform (FT) surrogates~\cite{Theiler1992}, which preserve each recording's linear correlation structure while enforcing Gaussian increments.
The LFP step-size distribution is mildly but robustly heavy-tailed (\cref{subfig:lfp_increments}): the median excess kurtosis of the increments is $\kappa = 0.35$ across sessions, whereas the surrogate value is indistinguishable from zero ($\Delta\kappa = 0.35$, $p < 10^{-20}$, one-sided sign test across the $69$ sessions).
Layer 2/3 LFP therefore departs from the Gaussian--Markovian limit in two directions at once---heavy-tailed non-Gaussianity and non-Markovian long-range memory---whose competing superdiffusive and subdiffusive contributions net to the superdiffusion seen in the MAD while producing the long-range temporal correlations seen in the PSD.

Neuronal spikes provide an independent measure of long-range temporal dependence.
We quantify spike-count variability using the Fano factor, $F(\tau) = \operatorname{Var}[N(\tau)]/\operatorname{E}[N(\tau)]\,,$ where $N(\tau)$ is the number of spikes in a window of duration $\tau$.
A homogeneous Poisson process has $F = 1$ at every window duration, whereas temporally correlated firing-rate fluctuations produce power-law growth, $F(\tau) \propto \tau^c$, with a positive `variability exponent' $c$ at long windows~\cite{Munn2020}.
In layer 2/3 of \texttt{VISp}, spike counts become super-Poissonian beyond windows of $\sim$\SI{0.01}{\second} and then grow as a power law with $c = 0.19$ (95\% CI: $[0.18, 0.21]$ across sessions; \cref{subfig:empirical_fano_factor}), corroborating the broadband temporal dependence of the LFP with a statistic that can be compared directly against the circuit model below.

Together, the LFP and spiking results identify a regime in which non-Gaussian superdiffusive fluctuations coexist with long-range temporal dependence, scale-dependent super-Poissonian spike variability, and oscillations.
We identify this joint regime as the AF state, since the non-Gaussian fluctuations and long-range temporal dependence are naturally represented by spatial and temporal fractional dynamics.
We observe similar dynamics in an independent dataset with longer epochs of spontaneous activity~\cite{Siegle2021} (\cref{fig:visual_coding}).
The values reported here characterize superficial \texttt{VISp}; related anomalous signatures appear across cortical depths and visual areas, so the AF state describes a family of cortical working points rather than a single universal set of exponents.
We next turn to a biophysical circuit model, in which the same signatures can be traced to recurrent interactions and adaptation, then to an effective theory that separates heavy-tailed excursions from temporal memory while retaining an oscillatory component (\nameref{sec:effective_theory}), and finally to the systematic variation of these exponents across the visual hierarchy (\nameref{sec:hierarchical_variation}).

\subsection{A cortical circuit near a localization transition exhibits adaptive fractional dynamics}

The Neuropixels recordings establish the AF state at the population and spiking levels but do not elucidate the underlying circuit mechanisms that generate superdiffusion, long-range memory, and oscillations.
In particular, the LFP provides an indirect measure of the synaptic drive to individual neurons, although related anomalous signatures have been reported in single-neuron subthreshold fluctuations~\cite{DeWeese2006,Johnson2019}.
We thus examine a biophysical spiking network in which both synaptic currents and membrane potentials are directly accessible, asking whether the joint signatures of non-Gaussian excursions, temporal dependence, and oscillations can emerge from collective circuit dynamics while individual neurons remain fluctuation-driven.

The model~\cite{Qi2022a} comprises spatially embedded excitatory and inhibitory neurons with distance-dependent recurrent coupling~\cite{Levy2012}, balanced excitation and inhibition~\cite{Okun2008}, and spike-frequency adaptation in the excitatory population; we provide full details in \nameref{sec:methods}.
Adaptation supports a slow $\theta$ oscillation in the local population activity, transiently suppressing neuronal firing after periods of $\gamma$ bursting generated by the PING mechanism~\cite{Qi2022a}.
Local coupling between the spatially embedded neurons gives rise to localized activity packets, whose stability depends on the effective balance between excitation and inhibition~\cite{Gu2019,Chen2022b}.
We examine a working point near the transition between a strongly `localized' regime, in which coherent wave packets propagate smoothly through the network, and a `delocalized' regime, in which spiking becomes spatially diffuse and weakly correlated.
Near this transition, weakly localized patterns wander anomalously across space (\cref{subfig:schematic_circuit}, animated in \cref{video:input_field}), and previous work has shown that the model reproduces diverse features of visual cortical dynamics in this regime~\cite{Gu2019,Chen2019,Chen2022b,Qi2022a,Ni2024}.
Importantly, unlike the experimental recordings, the model provides direct access to both the synaptic currents arriving at individual neurons and the resulting subthreshold membrane-potential dynamics (\cref{subfig:membrane_potential,subfig:potential_distribution}).

Individual neurons remain fluctuation-driven despite this structured population activity: the mean membrane potential is \SI{-59}{\milli\volt}, approximately \SI{9}{\milli\volt} below the \SI{-50}{\milli\volt} spiking threshold, so that fluctuations produce intermittent threshold crossings and sparse firing (\cref{subfig:membrane_potential,subfig:potential_distribution}), consistent with single-neuron recordings in vivo~\cite{Ahmadian2021}.
Synaptic input currents exhibit large, irregular `bumps' (\cref{subfig:input_current}), and their short-timescale increments have a strongly heavy-tailed, near-power-law distribution spanning approximately two decades of amplitude (\cref{subfig:input_distribution}); because deterministic drift vanishes relative to the stochastic drive over sufficiently short increments, these step-size statistics are a direct signature of the stochastic drive.
The heavy tails of the recurrent input are better captured by a L\'evy distribution, giving substantially greater probability of large excursions than expected under Gaussian fluctuations~\cite{Wardak2021,Becker2024}.
Sparse, fluctuation-driven firing therefore coexists with larger input excursions than expected from the Gaussian approximation of conventional mean-field models~\cite{Brunel2000,Volo2019}.

The synaptic currents also reproduce the joint superdiffusion and shallow spectra identified in the experimental data.
The input currents are superdiffusive at short timescales: their MAD scales with a median diffusion exponent of $a = 0.599$ (95\% CI: $[0.597, 0.601]$ across neurons) for lags up to \SI{8}{\milli\second} (\cref{subfig:empirical_mad}).
This superdiffusion depends on the synchronous arrival of presynaptic spikes, a collective property of the circuit rather than of individual spike trains: after independently shifting each presynaptic spike train, the diffusion exponent of the input falls from $0.61 \pm 0.01$ to $0.53 \pm 0.01$ (mean $\pm$ SD over $10$ network realizations).
At the same time, the input-current spectrum exhibits broadband temporal dependence, with a spectral exponent of $b = -1.735$ (95\% CI: $[-1.741, -1.729]$ across neurons) alongside $\theta$- and $\gamma$-range peaks at approximately \SI{4}{\hertz} and \SIrange{40}{120}{\hertz} (\cref{subfig:empirical_psd}).
These input dynamics propagate to neuronal output: the spike-count Fano factor of excitatory neurons is approximately Poissonian at short windows ($\tau < \SI{0.01}{\second}$), increases as a power law over intermediate windows ($c = 0.28$, fitted scaling band \SIrange{10}{100}{\milli\second}), and saturates at longer timescales (\cref{subfig:empirical_fano_factor}).
Although the relationship between input fluctuations (studied in the circuit model) and the local field potential (studied in the experimental data) is complex, the Fano factor scaling of spike counts corroborates the presence of long-range memory across both the circuit model and Neuropixels recordings.

Thus, the circuit recapitulates the key dynamical signatures observed experimentally: short-timescale superdiffusion ($a > 0.5$), broadband temporal dependence ($-2 < b < 0$), coexisting $\theta$- and $\gamma$-range rhythms, and scale-dependent spike-count variability ($c > 0$).
Crucially, these signatures do not arise from arbitrary parameter tuning: for biophysically realistic levels of adaptation, PING oscillations, time constants, and local connectivity, the anomalous dynamics emerge in the transition regime, which best captures diverse experimental observations~\cite{Gu2019,Qi2022a,Ni2024}.
Together, the experimental and circuit results identify the AF state as a unifying cortical working regime in which the driving fluctuations are simultaneously non-Gaussian, history-dependent, and oscillatory; properties that conventional mean-field theories capture only individually~\cite{Brunel2000,Volo2019}.

\subsection{A minimal effective theory links superdiffusion, memory, and oscillations}
\label{sec:effective_theory}

The experimental and circuit results demonstrate three defining components of the AF working regime: heavy-tailed superdiffusive excursions, long-range temporal dependence, and oscillations.
Conventional mean-field models provide limited control over these components; their stochastic drive is typically approximated as white Gaussian noise, fixing the increment statistics and temporal dependence to the classical diffusive limit (with a diffusion exponent of $0.5$ and a spectral exponent of $-2$), while oscillations are introduced through additional deterministic variables~\cite{Brunel2000,Volo2019}.
We therefore develop a minimal effective theory in which the prevalence of heavy-tailed excursions, the strength of temporal memory, and the oscillatory component can be separately tuned.
Our formulation provides a parsimonious description of the joint superdiffusion, broadband temporal dependence, and oscillations observed in both the mouse brain and the circuit model; here, `effective' denotes an empirical mean-field description constrained by the statistics of the synaptic input, rather than a self-consistent analytical reduction of the underlying circuit.

Fractional calculus provides an elegant formalism for capturing long-range and non-Gaussian effects that can give rise to anomalous diffusion.
Broadly, the operators of fractional calculus generalize classical derivatives to non-integer orders by incorporating power-law heavy-tailed kernels.
The fractional spatial derivative introduces step-size distributions with power-law tails, and has been used to model anomalous superdiffusion in membrane potential dynamics~\cite{Wardak2021}.
On the other hand, the fractional temporal derivative introduces long-range temporal dependence, and has been used to model multiscale adaptation processes in single neurons~\cite{Lundstrom2008,Lundstrom2023}.

We build on the fractional neural sampling (FNS) framework of \cite{Qi2022a}, which proposes a Langevin-like process combining L\'evy-driven motion with an auxiliary momentum variable, and which was originally used to model the motion of the localized activity pattern in the biophysical circuit.
Here, we aim to model the synaptic input dynamics of the circuit, capturing long-range temporal dependencies alongside L\'evy superdiffusion.
We therefore extend FNS by introducing a fractional temporal derivative, giving a fully bi-fractional model---which we term bi-fractional neural sampling (bFNS)---defined on the one-dimensional position $x$ and auxiliary momentum $p$:
\begin{equation}
	\label{eq:bifractional_neural_sampling}
	\begin{aligned}
		{}^C D_t^\beta x &= -\eta \nabla \tilde{V}_\alpha + \gamma p + \eta^{\frac{1}{\alpha}} \xi_{\alpha,\beta}\,,\\
		\frac{dp}{dt} &= -\gamma \nabla \tilde{V}_\alpha \,,
	\end{aligned}
\end{equation}
where $\alpha \in (1, 2]$ is the spatial fractional order, ${}^C D_t^\beta$ is the Caputo fractional derivative of order $\beta \in (0, 1]$ [see \cref{eq:caputo_derivative}], $\eta$ sets the stochastic timescale, $\gamma$ couples $x$ to the auxiliary oscillatory momentum, and $\xi_{\alpha,\beta}$ is a linear fractional stable drive.
The effective potential $\tilde{V}_\alpha$ is constructed so that the process samples a specified target distribution $\pi(x)$, which we verify numerically in \cref{fig:theoretical_model_unimodal,fig:theoretical_model_bimodal}; full definitions and the numerical scheme are given in \cref{sec:theoretical_model_sup}.

The three components of bFNS have distinct effects, illustrated in \cref{fig:theoretical_model} through their influence on deterministic (`drift') and stochastic (`diffusion') dynamics.
The spatial fractional order controls the non-Gaussian content of the drive: lowering $\alpha$ below $2$ gives the driving noise power-law tails and admits occasional large increments (\cref{subfig:space_fractional_diffusion}), while reshaping the effective potential through which the process evolves (\cref{subfig:space_fractional_drift}).
The temporal fractional order introduces history dependence: for $\beta < 1$ the Caputo derivative weights past increments through a slowly decaying kernel and correlates the driving noise in time (\nameref{sec:methods}), so that relaxation becomes power-law rather than exponential (\cref{subfig:time_fractional_drift}) and the unconfined power spectrum grows shallower as $\beta$ falls (\cref{subfig:time_fractional_diffusion}).
Memory therefore acts on the broadband spectrum by opposing the growth of fluctuations with time lag, a distinct effect to $\alpha$.
The momentum coupling supplies the oscillatory degree of freedom: through \cref{eq:bifractional_neural_sampling} the momentum $p$ accumulates the restoring force $-\gamma\nabla \tilde{V}_\alpha$ that feeds back into the evolution of $x$, producing oscillatory motion around local minima of the potential (sample time series are shown in \cref{subfig:bFNS_traces}).
The model incorporates heavy-tailed excursions, temporal memory, and oscillations through separate components, although their effects on the measured statistics interact: $\alpha$ and $\beta$ jointly determine the diffusion exponent, and the self-similarity of the linear fractional stable drive is linked to both orders (\nameref{sec:methods}).

Choosing, for example, $\alpha = 1.5$, $\beta = 0.85$, $\gamma = 0.03$, and $\eta = 0.01$ gives a model with the mild superdiffusion, long-range memory, and oscillations that characterize the AF regime (\cref{fig:theoretical_model}).
\Cref{subfig:diffusion_exponent} shows that the model exhibits superdiffusive dynamics across all time lags when exploring a flat, unconfined potential, and at short time lags when confined on a unimodal potential.
\Cref{subfig:spectral_exponent} shows that the model exhibits a power-law PSD in the unconfined case, but in a unimodal potential shows a peaked PSD with a power-law high-frequency tail, reflecting local oscillations.

Together, the three parameters allow bFNS to unify several established stochastic descriptions of neural dynamics:
\begin{equation}
	\label{eq:bfns_limits}
	\text{bFNS}(\alpha, \beta, \gamma) \rightarrow
	\begin{cases}
		\enspace\text{Gaussian--Markovian diffusion}\,, & \alpha = 2,\ \beta = 1,\ \gamma = 0\,,\\
		\enspace\text{Oscillatory Gaussian dynamics}\,, & \alpha = 2,\ \beta = 1,\ \gamma > 0\,,\\
		\enspace\text{L\'evy-driven superdiffusion}\,, & \alpha < 2,\ \beta = 1,\ \gamma = 0\,,\\
		\enspace\text{Gaussian dynamics with power-law memory}\,, & \alpha = 2,\ \beta < 1,\ \gamma = 0\,,\\
		\enspace\text{Full bi-fractional oscillatory regime}\,, & \alpha < 2,\ \beta < 1,\ \gamma > 0\,.
	\end{cases}
\end{equation}
With $\alpha = 2$, $\beta = 1$, and $\gamma = 0$, bFNS reduces to the white Gaussian drive assumed by classical mean-field approximations~\cite{Brunel2000}, and setting $\gamma > 0$ adds an oscillatory component without changing the Gaussian input statistics.
Lowering $\alpha$ alone recovers a L\'evy-driven process with heavy-tailed increments but no long-range memory~\cite{Wardak2021}, whereas lowering only $\beta$ produces a temporally fractional process with power-law memory but light-tailed fluctuations~\cite{Li2017b}.

By varying $\alpha$, $\beta$, and $\gamma$, the bFNS model can capture a wide range of diffusion and spectral exponents (\cref{subfig:diffusion_exponent_sweep,subfig:spectral_exponent_sweep}).
For unconfined motion, the diffusion exponent follows the approximate relation $a = 1 - \alpha/2 + \beta/2$ (within $0.1$ of the fitted exponents over $90\%$ of the valid region of \cref{subfig:diffusion_exponent_sweep}, and within $0.15$ everywhere), meaning that the fractional spatial derivative creates superdiffusion (decreasing $\alpha$ increases $a$) and the fractional temporal derivative creates a competing subdiffusion (decreasing $\beta$ decreases $a$).
This empirical relation describes the position process $x$ and differs from the self-similarity exponent $H = 1/2 + 1/\alpha - \beta/2$ chosen for the integrated noise process (\nameref{sec:methods}); unlike $a$, $H$ decreases with $\beta$.
On the other hand, the spectral exponent $b$ is primarily controlled by $\beta$, with only a weak dependence on $\alpha$ at small values of $\beta$ (\cref{subfig:spectral_exponent_sweep}).
This partial decoupling of the diffusion and spectral exponents allows the bi-fractional model to tune the degree of superdiffusion and long-range temporal dependence separately, capturing joint anomalous effects that are not possible with previous L\'evy process or fractional Brownian motion models.

In summary, $\alpha$ primarily controls the non-Gaussian contribution to the diffusion exponent $a$; $\beta$ controls long-range memory and hence the spectral exponent $b$, while also opposing superdiffusion; and $\gamma$ controls the oscillatory component.
Together, these parameters place Gaussian, L\'evy-like, memory-dependent, and oscillatory regimes within a common dynamical framework.
By converting the dynamical signatures observed in the recordings and circuit model into interpretable degrees of freedom, bFNS provides the theoretical bridge needed to ask how cortical working regimes vary across neurons, circuits, and cortical areas.

\subsection{An empirical bi-fractional mean field reproduces the adaptive fractional state}

Having established that bFNS captures the dynamical coordinates of the AF state, we next ask how these anomalous input statistics shape the activity of individual neurons.
Specifically, can an effective input process constrained by the synaptic statistics of the recurrent circuit generate the subthreshold and spiking signatures observed at the neuronal level?
To address this question, we use bFNS as an empirical mean field for the total synaptic input to a single fluctuation-driven neuron (\cref{subfig:mean_field_schematic}).
Unlike conventional mean-field reductions, which typically approximate the recurrent input as a white Gaussian drive~\cite{Brunel2000}, our construction retains the heavy-tailed fluctuations, long-range temporal memory, and oscillatory structure identified in the circuit.
Rather than deriving a self-consistent analytical reduction of the network, we use this framework to isolate how distinct statistical features of the synaptic drive shape membrane-potential and spiking dynamics.

We first constrain the stationary input distribution against the synaptic currents measured directly from the circuit model.
Fitting these currents with a stable distribution gives a representative target density with a tail index of $1.5$, skewness of $0$, scale of $0.14$, and location of $0.2$, which fixes the effective potential of the bFNS process (fitted parameter distributions across neurons are shown in \cref{sup:input_distribution_parameters}; see \cref{subsec:mean_field_model_sup}).
We then choose a representative dynamical parameter set ($\alpha = 1.5$, $\beta = 0.8$, $\eta = 0.03$, and $\gamma = 0.01$) that lies within the AF regime and yields physiologically realistic firing and dynamical statistics.
Driven by this bFNS input, the mean-field neuron fires sparsely at approximately \SI{8.8}{\hertz} while the mean membrane potential remains $\sim$\SI{12}{\milli\volt} below threshold (\cref{subfig:membrane_potential_mean_field,subfig:potential_distribution_mean_field}), recapitulating the fluctuation-driven regime of the full circuit.
The input itself retains the irregular, `bumpy' fluctuations and heavy-tailed step-size distribution characteristic of the circuit synaptic currents (\cref{subfig:input_current_mean_field,subfig:input_distribution_mean_field}).

We now seek to understand how fractional effects influence firing rates by systematically varying $\alpha$ and $\beta$ while fixing the other default parameters.
As shown in \cref{subfig:firing_rate_mean_field}, the increase in superdiffusion caused by reducing $\alpha$ generally leads to greater firing rates, as large input fluctuations more frequently drive the membrane potential across threshold.
The relationship between firing rate and $\beta$, however, is more complex.
We observe that decreasing $\beta$ first tends to increase firing rate---by maintaining elevated input levels for longer periods---before decreasing firing rate at very low $\beta$ values.
Thus, heavy-tailed excursions and temporal memory influence neuronal output through different mechanisms, producing an enhanced fluctuation-driven regime in which substantial firing arises without the mean membrane potential approaching threshold.


Finally, we examine how the fractional orders shape spike-time variability in the mean-field neuron, comparing the full model against restricted versions that correspond to limiting cases of \cref{eq:bfns_limits} (\cref{subfig:fano_factor_mean_field}).
With no long-range memory ($\beta = 1$), the Fano factor curve is flat and sits near unity, indicating Poisson-like spiking statistics.
With memory restored but momentum removed ($\beta < 1$, $\gamma = 0$), the time-fractional effects introduce a power-law scaling at long time windows, as observed in the experimental data.
Introducing momentum and oscillations ($\gamma > 0$) tempers the power-law scaling and flattens the Fano factor curve at long timescales, as observed in the circuit model.
That the circuit model exhibits a tempered power-law Fano factor, while the experimental data show a power law extending to windows of one second, may reflect heterogeneity in the rhythmicity of firing across real neural populations, or intracellular dynamics not captured by the circuit model.
Across the full parameter sweep (\cref{subfig:variability_exponent_mean_field}), the variability exponent $c$ climbs to $0.5$ as the fractional orders are varied, and turns negative near $\beta = 1$, where the Fano factor declines with window length in the Markovian sampling limit.

Together, these results establish bFNS as a cross-scale effective description of the AF state: with the target distribution fixed by the input statistics of the circuit, the mean-field neuron reproduces the subthreshold membrane-potential distribution, fluctuation-driven firing, and the scale dependence of spike-count variability observed in the circuit and experimental data.
Crucially, bFNS incorporates only the minimal degrees of freedom necessary to capture these anomalous scaling properties, alongside a single oscillatory component, making it a parsimonious approximation to input fluctuations.
The bFNS mean field thus provides a unified framework for understanding how the anomalous statistics observed at the synaptic level propagate to the subthreshold and spiking dynamics of individual neurons.

\subsection{Working regimes vary systematically across the visual hierarchy}
\label{sec:hierarchical_variation}

So far, we have treated the adaptive fractional state as a fixed working point, characterized by a pair of dynamical exponents in layer 2/3 of \texttt{VISp}.
A defining feature of the bFNS framework, however, is that the fractional orders $\alpha$ and $\beta$ move the diffusion and spectral exponents along distinct directions in the $(a, b)$ plane (\cref{fig:theoretical_model}).
This flexibility raises the possibility that cortical areas occupy different combinations of superdiffusive excursions and temporal memory, according to their local circuit organization and computational demands.
We therefore ask whether areas along the visual hierarchy differ primarily in a single characteristic timescale, as commonly proposed~\cite{Murray2014}, or instead occupy distinct joint working points that combine different degrees of superdiffusion and long-range temporal dependence.

Across the six visual areas (\cref{subfig:cortex_map}), the dynamical exponents measured in superficial layer 2/3 fall along a common axis in the $(a, b)$ plane, broadly ordered by anatomical hierarchy score~\cite{Harris2019d,Siegle2021} (\cref{subfig:hierarchy_l23}).
The primary visual area \texttt{VISp} is the most strongly superdiffusive, with a median diffusion exponent of $a = 0.62$ (95\% CI: $[0.60, 0.64]$ across sessions) and a spectral exponent of $b = -1.75$ (95\% CI: $[-1.79, -1.72]$).
By contrast, the highest area, \texttt{VISam}, approaches normal diffusion, with $a = 0.51$ (95\% CI: $[0.48, 0.54]$), while its spectrum is substantially shallower, with $b = -1.57$ (95\% CI: $[-1.60, -1.55]$), indicating stronger long-range temporal correlations.
This finding suggests that short-timescale fluctuations become less superdiffusive toward higher areas while broadband temporal dependence becomes stronger, shifting cortical activity along a joint dynamical axis from stronger rapid excursions toward greater temporal persistence.

We quantify this hierarchical organization across cortical layers by correlating each dynamical exponent with anatomical hierarchy score.
Within each session, we rank that session's areas against their hierarchy scores to obtain a single Kendall's $\tau$ per exponent and layer, and report the across-session median (\nameref{sec:methods}).
The hierarchical organization is strongest in superficial cortex (\cref{subfig:hierarchy_correlation}): in layer 2/3, the diffusion exponent decreases with hierarchy score ($\tau = -0.40$, 95\% CI: $[-0.47, -0.33]$) while the spectral exponent increases ($\tau = 0.60$, 95\% CI: $[0.47, 0.60]$; $p < 10^{-3}$), so the two exponents change in opposite directions along the hierarchy, with higher areas showing weaker superdiffusion but stronger long-range memory.
The joint gradient does not survive into deeper layers: the spectral exponent becomes uncorrelated by layer 6 ($\tau = -0.07$, $p = 0.9$), while the diffusion exponent weakens through layers 4 and 5 and then reverses sign, becoming positive in layer 6 ($\tau = 0.33$, $p < 10^{-3}$).
The variability exponent, measured from spikes rather than from the LFP, is positively correlated with hierarchy score at every layer, matching the sign of the spectral exponent ($\tau$ between $0.20$ and $0.33$); the correlation is significant in layers 2/3, 5, and 6 ($p < 10^{-3}$) but not in layer 4 ($\tau = 0.20$, $p = 0.05$).
The layer 2/3 pattern also replicates in the Allen Visual Coding cohort (\cref{fig:visual_coding}).
The joint, oppositely signed displacement of $a$ and $b$ that defines the hierarchical axis is therefore concentrated in superficial layers, consistent with the prominent role of mouse layer 2/3 in recurrent and interareal processing~\cite{Harris2019d,D’Souza2016}.

The bFNS effective theory allows us to interpret this dynamical gradient.
Changing the spatial fractional order $\alpha$ primarily alters the prevalence of large, non-Gaussian excursions, whereas changing the temporal fractional order $\beta$ alters the power-law memory and therefore the spectral exponent (\cref{subfig:diffusion_exponent_sweep,subfig:spectral_exponent_sweep}).
In the $(a, b)$ plane, the axis from \texttt{VISp} toward higher areas aligns predominantly with the direction of stronger time-fractional effects (lower $\beta$) rather than with a change in excursion statistics via $\alpha$ (\cref{subfig:hierarchy_l23}).
Lowering $\beta$ simultaneously reduces net diffusion and makes the spectral slope shallower, reproducing the direction of variation observed across the hierarchy.
The data therefore imply that the principal hierarchical change is an increasing influence of past activity on the present state, with the tempering of rapid superdiffusive excursions arising from the same time-fractional shift.

To determine which circuit mechanisms could move cortical dynamics along this hierarchical axis, we systematically vary pairs of biophysical parameters in the circuit model that alter the organization of recurrent activity, holding others at the default values used for \cref{fig:experimental_data}, then recompute the dynamical exponents of the synaptic input (see \nameref{sec:methods}).
Here we focus on the plane spanned by the synaptic I:E ratio $\delta$ and the excitatory adaptation strength $\Delta g_K$, which we find have the greatest influence on the dynamical exponents; \cref{subfig:circuit_diffusion_map,subfig:circuit_spectral_map} show the resulting exponents against $\delta$ for five values of $\Delta g_K$ spanning \SIrange{0.001}{0.005}{\micro\siemens}.
At the default working point ($\delta = 3.75$, $\Delta g_K = \SI{0.002}{\micro\siemens}$; \cref{tab:params}), the input dynamics are strongly superdiffusive, with $a = 0.610$ (95\% CI: $[0.599, 0.616]$ across network realizations) and $b = -1.755$ (95\% CI: $[-1.760, -1.749]$).
At higher $\delta$, strong relative inhibition drives the diffusion exponent toward a saturated value near $0.64$ (\cref{subfig:circuit_diffusion_map}): in this inhibition-dominated regime, firing is sparse and spatially disorganized, and the input to a neuron is dominated by rare, large synaptic events, yielding strongly superdiffusive input statistics.
The working point instead sits at slightly weaker inhibition, within the AF regime, in which activity packets coalesce and propagate through the network.
As $\delta$ is reduced through the AF regime, the diffusion exponent declines and ultimately falls below $0.5$ during the transition to the excitation-dominated regime, in which activity packets freeze and neurons within a packet fire at saturated rates.
Meanwhile, the spectral exponent becomes progressively less negative, rising from $\approx -1.87$ in the strongly inhibited, strongly adapting corner of the plane toward $\approx -1.07$ as $\delta$ and $\Delta g_K$ fall (\cref{subfig:circuit_spectral_map}).
Lowering the effective inhibition therefore moves the circuit in the same direction observed from lower to higher visual areas, reducing large excursions by enhancing temporal persistence.

Within the transition regime, changing the circuit balance alters both how strongly activity moves through state space and how long previous configurations continue to influence it.
Nevertheless, the synaptic I:E ratio provides a parsimonious circuit-level axis whose displacement in the $(a, b)$ plane closely matches the principal direction of hierarchical variation observed in the recordings (\cref{subfig:hierarchy_l23}).
This agreement between experimental data, the circuit model, and the effective theory establishes a common interpretation of the cortical hierarchy.
Increasing hierarchical position is associated with decreasing $a$ and increasing $b$; in the circuit, reducing the effective dominance of inhibition produces the same joint displacement; and in bFNS, strengthening time-fractional memory by lowering $\beta$ moves the dynamics in the same direction.
This shared dynamical axis links circuit properties to the rapid superdiffusive excursions and temporal memory.
Lower areas occupy a working point with stronger superdiffusive fluctuations, enabling rapid responses and large transitions following changes in sensory input; higher areas occupy a working point with fewer large excursions and stronger temporal memory, favoring the persistence and integration of information across time.
The $(a, b)$ plane therefore provides a compact and experimentally accessible map of cortical working regimes, locating areas by the statistics of their fluctuations and connecting those positions both to the fractional orders of the effective theory and to a circuit mechanism that can be tested directly.


\section{Discussion}
\label{sec:discussion}

By combining Neuropixels recordings from mouse visual cortex, mechanistic circuit modeling, and a generalized mean-field theory, we have developed a unifying framework that captures and explains a diverse set of dynamical features of the cortical working regime: irregular fluctuation-driven spiking, heavy-tailed non-Gaussian input fluctuations, long-range temporal dependence, super-Poissonian spike-count variability, and coexisting $\theta$ and $\gamma$ oscillations.
Together, these features define the AF state, unifying departures from the Gaussian--Markovian limit that have previously been studied in isolation.
The AF state provides a principled account of how cortical working regimes are organized along the visual hierarchy and of the neurophysiological mechanisms underlying this organization, offering new insight into cortical computation and yielding testable predictions.

The AF state extends established accounts of cortical dynamics.
Classical balanced-state theories explain irregular, low-rate firing through the dynamic cancellation of excitation and inhibition~\cite{Vreeswijk1996,Brunel2000}, and subsequent extensions have incorporated oscillations~\cite{Brunel1999,Brunel2003}, temporal correlations~\cite{MorenoBote2004,Fourcaud2002}, and heavy-tailed fluctuations~\cite{Wardak2021,Becker2024}, while largely relaxing the Gaussian and Markovian assumptions individually.
As we have shown, neither heavy-tailed non-Gaussianity nor non-Markovian memory alone accounts for the superdiffusive excursions, shallow power spectra, and super-Poissonian spiking variability that appear together in the experimental data and circuit simulations; only a jointly non-Gaussian and non-Markovian description accounts for all three, placing the recordings, the circuit model, and the mean-field theory in the same $(a, b)$ plane.

Our results also reframe the dynamical organization of the cortical hierarchy.
A prominent view holds that cortical areas are ordered along a hierarchy of intrinsic timescales, with progressively longer integration windows toward higher-order cortex~\cite{Murray2014,Rudelt2024, Shi2025}.
Our findings support the underlying idea of increasing temporal integration, but show that a single characteristic timescale is an incomplete summary: from \texttt{VISp} to \texttt{VISam} the diffusion exponent falls from $0.62$ to $0.51$ while the spectral exponent rises from $-1.75$ to $-1.57$, so that the two coordinates move in opposite directions along a single axis in the $(a, b)$ plane, ordered by anatomical hierarchy score.
Thus, higher visual areas exchange superdiffusive excursions for temporal persistence.

This hierarchical variation is also specific to superficial cortex.
Both gradients are strongest in layer 2/3, consistent with the prominent role of mouse superficial layers in recurrent and interareal processing~\cite{Harris2019d,D’Souza2016}, whereas by layer 6 the correlation of the diffusion exponent has reversed sign and that of the spectral exponent has vanished; the joint axis is a property of superficial circuits rather than a global feature of an area.
The variability exponent, which is measured from spikes rather than from the LFP and so from a different signal in the same recordings, nonetheless increases with hierarchy score at every layer (significantly in all but layer 4), corroborating the strengthening of long-range temporal dependence toward higher areas.
The AF framework therefore generalizes the classical hierarchy of timescales to a hierarchy of dynamical working regimes with joint space- and time-fractional properties.

Which circuit mechanisms organize this hierarchy of working regimes?
In our circuit model a single control parameter, the synaptic I:E ratio $\delta$, displaces both exponents together along a direction closely aligned with the hierarchical axis measured experimentally: reducing the relative dominance of inhibition weakens large-excursion dynamics while increasing temporal persistence.
We attribute this displacement to the model's proximity to the transition between localized and delocalized activity.
Under strong relative inhibition, firing is sparse and spatially disorganized, so a neuron's input is dominated by rare, large synaptic events and its statistics are strongly superdiffusive.
As inhibition weakens, activity packets coalesce and propagate coherently, so that the input a neuron receives depends increasingly on the recent trajectory of the packet, converting rapid excursions into temporal persistence.
Since higher visual areas lie toward the weak-superdiffusion, high-persistence end of this axis, the effect of ascending the hierarchy is reproduced in the circuit model by raising relative excitation to approach the transition point.
This direction is consistent with known anatomical gradients: parvalbumin-expressing interneurons, which supply fast inhibition, are relatively more abundant in lower visual areas~\cite{Kim2017}, and the inhibitory contribution to interareal connectivity decreases toward higher levels of the hierarchy~\cite{D’Souza2016}, with both gradients strongest in layer 2/3, where the hierarchical organization of the exponents is clearest.
The correspondence should not be read as uniquely identifying E:I balance from the measured exponents: synaptic kinetics, connectivity, adaptation, and other circuit properties can produce related displacements, and anatomical measures of inhibitory cell density do not directly specify the effective synaptic balance experienced by an individual neuron.
Rather, the agreement among cortical dynamics, circuit simulations, and anatomical gradients identifies effective circuit balance as one axis along which cortical working regimes can be organized, consistent with broader evidence that graded changes in excitation and inhibition produce qualitatively different dynamical regimes across cortex~\cite{Wang2020}.

The two coordinates of the AF state confer distinct computational advantages, which suggests why cortex might distribute itself along this axis.
Sampling-based theories propose that neural variability and noise can represent uncertainty and support probabilistic computation~\cite{Orban2016}, and previous work showed that the heavy-tailed motion of activity patterns allows cortical activity to efficiently sample multimodal probability landscapes~\cite{Qi2022a,Chen2022b}, providing a mechanism for flexible neural computation.
At the level of individual neurons studied here, large input excursions permit threshold crossings while the mean membrane potential remains far below threshold, enhancing the fluctuation-driven regime.
Long-range memory provides a complementary advantage, coupling the present to a slowly decaying history of past activity to stabilize local trajectories, extend `fading memory'~\cite{Yiling2024}, and produce the scale-dependent spike-count variability observed experimentally.
The AF state therefore combines two properties usually considered separately: large excursions expand the range of states that can be rapidly accessed, whereas long-range memory increases the persistence with which local states are maintained and integrated.

This competition between superdiffusion and subdiffusion suggests a computational interpretation of the hierarchical gradient, resembling a trade-off between exploration and exploitation.
Primary visual cortex is strongly superdiffusive, with large excursions that may facilitate rapid reconfiguration to changing sensory input.
Higher visual areas progressively temper the degree of superdiffusion by strengthening temporal memory, potentially favoring the accumulation and maintenance of information over extended periods.
Thus the dynamical hierarchy we observe through the diffusion exponent and spectral exponent generates the testable hypothesis that the optimal balance between excursion and memory should depend on the computational demands placed on a cortical area, and should shift with behavioral state, sensory context, and task demands.
This formulation also suggests that flexibility and stability can coexist within a single working regime and be continuously tuned by the circuit properties, such as E:I balance, that control the degree of long-range memory.

Beyond the diffusive properties of cortical activity, the AF state also accommodates the coexisting oscillatory dynamics that are thought to temporally organize local computation and interareal communication~\cite{Fries2015,Buzsaki2023}.
In the AF state, oscillations coexist with heavy-tailed excursions and long-range memory, captured parsimoniously by the bFNS model.
Future work may incorporate multiple oscillatory components to capture the nested $\theta$--$\gamma$ interactions of the circuit model~\cite{Chen2022b} and of experimental data~\cite{Harris2026}.
Further directions for theoretical development include a bi-fractional Fokker--Planck derivation for an analytic description of the self-similarity exponent and firing-rate variability, as well as a full spatially extended stochastic neural field model that could connect the bi-fractional modeled inputs to the wandering activity packets of the circuit model~\cite{Bressloff2011}.

Thus our study of the adaptive fractional state generates testable predictions at multiple levels.
LFP and spiking signatures are only indirect reflections of synaptic inputs, so we predict that intracellular recordings across the visual hierarchy should reveal synaptic inputs whose long-range temporal persistence strengthens toward higher areas while their heavy-tailed step statistics are comparatively preserved, most prominently in layer 2/3.
In summary, the adaptive fractional state encapsulates both the oscillations and the anomalous diffusive dynamics observed in cortical recordings and reproduced by a biophysical circuit model.
Our bi-fractional mean-field framework shows how these properties can arise from non-Gaussian fluctuations and long-range memory in subthreshold activity, and supplies a two-dimensional, experimentally accessible coordinate system in which cortical areas, layers, and circuit manipulations can be placed on the same axes.
The framework thereby bridges circuit mechanisms and hierarchical neural dynamics, reframing the cortical hierarchy as a trajectory through a space of dynamical working regimes.

\section{Resource availability}

\subsection{Lead contact}
Further information and requests for resources should be directed to and will be fulfilled by the lead contact, Pulin Gong (\href{mailto:pulin.gong@sydney.edu.au}{pulin.gong@sydney.edu.au}).

\subsection{Materials availability}
This study did not generate new materials or reagents.

\subsection{Data and code availability}
This study analyzes publicly available data.
The Visual Behavior Neuropixels dataset~\cite{Bennett2025} and the Visual Coding Neuropixels dataset~\cite{Siegle2021} are distributed by the Allen Institute for Brain Science and accessed via the AllenSDK (\href{https://allensdk.readthedocs.io/en/latest/}{https://allensdk.readthedocs.io/en/latest/}).
Processed data for re-plotting all figures are available on figshare, along with source data files for all plotted data and statistical results~\cite{Harris2026l}.

All analysis code for this study, including scripts that generate all figures from the raw data, is publicly available.
A manuscript repository is available at \href{https://github.com/brendanjohnharris/WorkingRegime.jl}{WorkingRegime.jl}~\cite{Harris2026m}, which includes instructions on how to reproduce all analyses and re-plot all figures from the deposited processed data.
This analysis repository draws on related projects: the spiking circuit model is implemented in \href{https://github.com/brendanjohnharris/Dewdrop.jl}{Dewdrop.jl}~\cite{Harris2026j}, the bi-fractional neural sampling model in \href{https://github.com/brendanjohnharris/FractionalNeuralSampling.jl}{FractionalNeuralSampling.jl}~\cite{Harris2026i}, and the MAPPLE estimator in \href{https://github.com/brendanjohnharris/TimeseriesTools.jl}{TimeseriesTools.jl}~\cite{Harris2026h}.

Any additional information required to reanalyze the data reported in this paper is available from the lead contact upon request.

\section{Acknowledgments}
This work was supported by the Australian Research Council (grant no.~DP160104316, P.G.).

\section{Author contributions}
B.H. and P.G. conceived and designed the study. B.H. performed the formal analysis and prepared the figures, under the supervision of P.G. B.H. and P.G. drafted, reviewed, and edited the manuscript.


\section{Declaration of interests}
The authors declare no competing interests.


\section{Supplemental information}
Document S1. Figures S1--S4.

Video S1. Wandering of the localized activity pattern in the circuit model, related to \cref{subfig:schematic_circuit}. MP4.


\clearpage
\section{Figures}

\begin{figure*}[!ht]
	\centering
	\includegraphics[width=\textwidth]{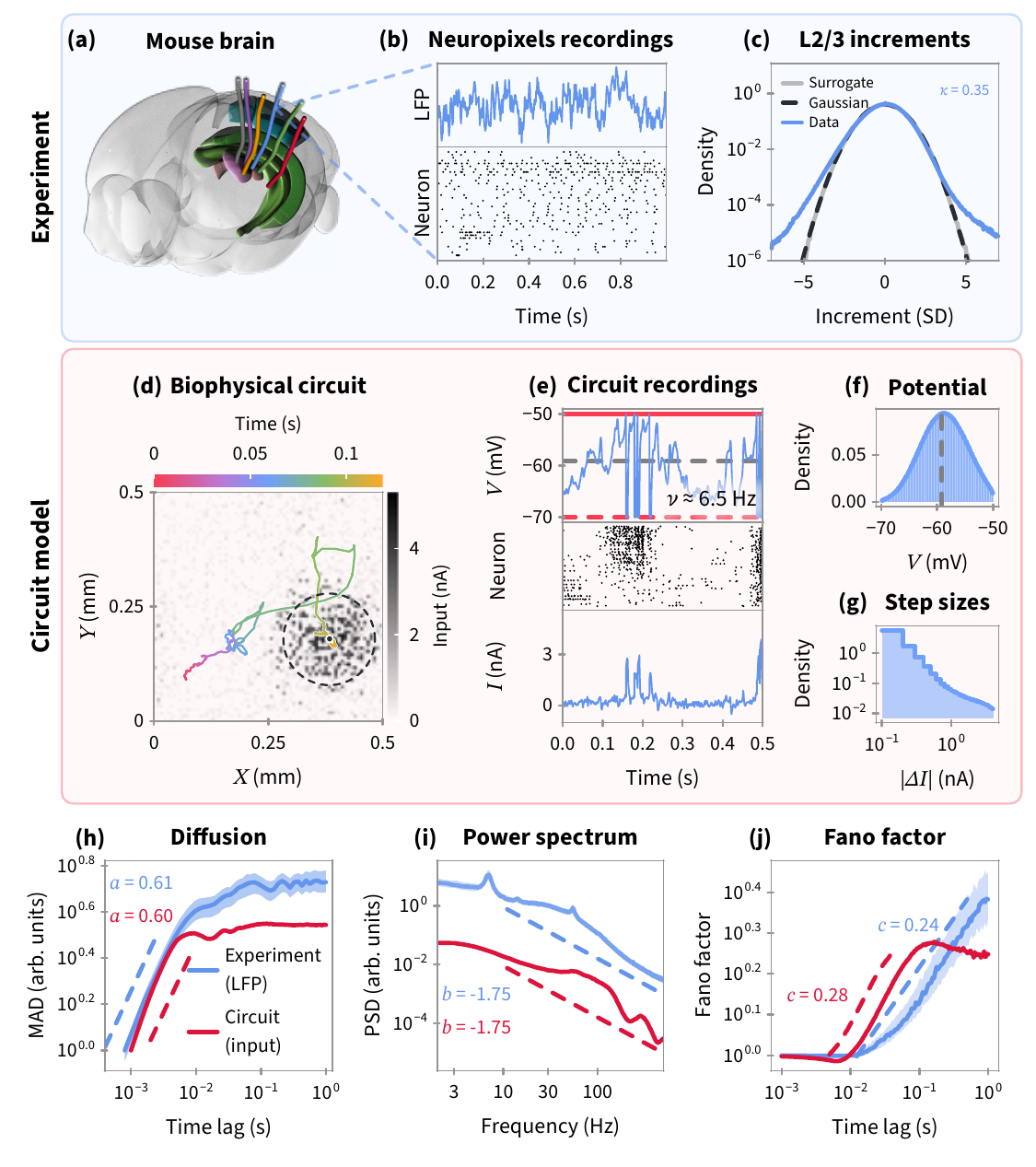}
	{
		\phantomsubcaption\label{subfig:schematic_experiment}
		\phantomsubcaption\label{subfig:brain_trace}
		\phantomsubcaption\label{subfig:lfp_increments}
		\phantomsubcaption\label{subfig:schematic_circuit}
		\phantomsubcaption\label{subfig:membrane_potential}\label{subfig:input_current}
		\phantomsubcaption\label{subfig:potential_distribution}
		\phantomsubcaption\label{subfig:input_distribution}
		\phantomsubcaption\label{subfig:empirical_mad}
		\phantomsubcaption\label{subfig:empirical_psd}
		\phantomsubcaption\label{subfig:empirical_fano_factor}
	}
	\caption{Cortical activity combines superdiffusion, long-range memory, and oscillations.
	\newsubref{subfig:schematic_experiment} Schematic of the mouse brain and Neuropixels probes.
	\newsubref{subfig:brain_trace} Example local field potential (LFP; upper) and spike-time (lower) recordings from layer 2/3 of the primary visual area (\texttt{VISp}) during gray-screen presentation. \\[0.5em](continued)
	}
	\end{figure*}
	\begin{figure*}
	\ContinuedFloat
	\centering
	\caption{
	\newsubref{subfig:lfp_increments} Distribution of LFP increments pooled across layer 2/3 channels (blue), each channel standardized by its own standard deviation, against a Gaussian (dashed black) and a phase-randomized (FT) surrogate that preserves each channel's power spectrum but enforces Gaussian increments (thin gray line).
	The annotated $\kappa$ is the median across sessions of the median excess kurtosis over channels; the surrogate value is indistinguishable from zero, so the heavy tails are not explained by the linear correlation structure.
	\newsubref{subfig:schematic_circuit} Snapshot of the input current in the circuit model, showing a localized activity pattern that wanders across space.
	The pattern's center-of-mass trajectory over the preceding window is shown as a colored line, and the dashed circle marks the neuronal patch analyzed in \subref{subfig:membrane_potential}.
	\newsubref{subfig:membrane_potential} Membrane potential of the excitatory neuron at the center of the patch in \subref{subfig:schematic_circuit} (upper), the spike raster of the surrounding patch population (middle), and the same neuron's synaptic input current, showing large, `bumpy' fluctuations (lower).
	The threshold and reset potentials are marked by solid and dashed red horizontal lines, respectively, the mean potential by a dashed gray line, and the neuron's median firing rate $\nu$ is annotated.
	\newsubref{subfig:potential_distribution} Distribution of membrane potentials across all excitatory neurons in the circuit model, showing a mean significantly below the spiking threshold.
	\newsubref{subfig:input_distribution} Step-size distribution for the synaptic input currents across all excitatory neurons in the circuit model, shown on log-log axes to highlight heavy, near-power-law tails.
	\newsubref{subfig:empirical_mad} Mean absolute deviation (MAD) of LFP fluctuations in layer 2/3 of the primary visual area (\texttt{VISp}, spontaneous activity; blue) and synaptic input currents in the circuit model (red), both showing a superdiffusive component with diffusion exponent $>0.5$ at short time lags.
	Dashed lines show power-law fits to the drawn median curves over the fitted band, with the diffusion exponent $a$ annotated; the curves are normalized in log-log coordinates, so only their slopes are comparable.
	For experimental data, the solid line and shaded band indicate the across-session median and IQR of within-session median values.
	\newsubref{subfig:empirical_psd} Power spectral density (PSD) of LFP fluctuations in \texttt{VISp} (blue) and synaptic input currents in the circuit model (red), both showing a broadband component with spectral exponent $>-2$, indicating long-range temporal dependence, alongside $\theta$ and $\gamma$ oscillatory peaks (at approximately \SI{7}{\hertz} and \SI{55}{\hertz} in the LFP).
	Dashed lines show the aperiodic components of a MAPPLE fit to each drawn curve, with the high-frequency exponent annotated.
	\newsubref{subfig:empirical_fano_factor} Fano factor of neuronal spikes in layer 2/3 of \texttt{VISp} (blue) and excitatory neurons in the circuit model (red), both showing super-Poissonian values ($>1$) and power-law scaling over intermediate time windows.
	Dashed lines, offset to the left for visibility, mark the scaling band fitted to each curve, and the variability exponent $c$ of each fit is annotated.
	The experimental fit spans only the windows drawn here, up to \SI{1}{\second}.
	The per-session fits quoted in the text extend to windows of \SI{10}{\second}, over which the rise of the Fano factor slows, and so give a smaller exponent (\nameref{sec:methods}).
	}
	\label{fig:experimental_data}
\end{figure*}
\clearpage
\begin{figure*}[htbp]
	\centering
	\includegraphics[width=\textwidth]{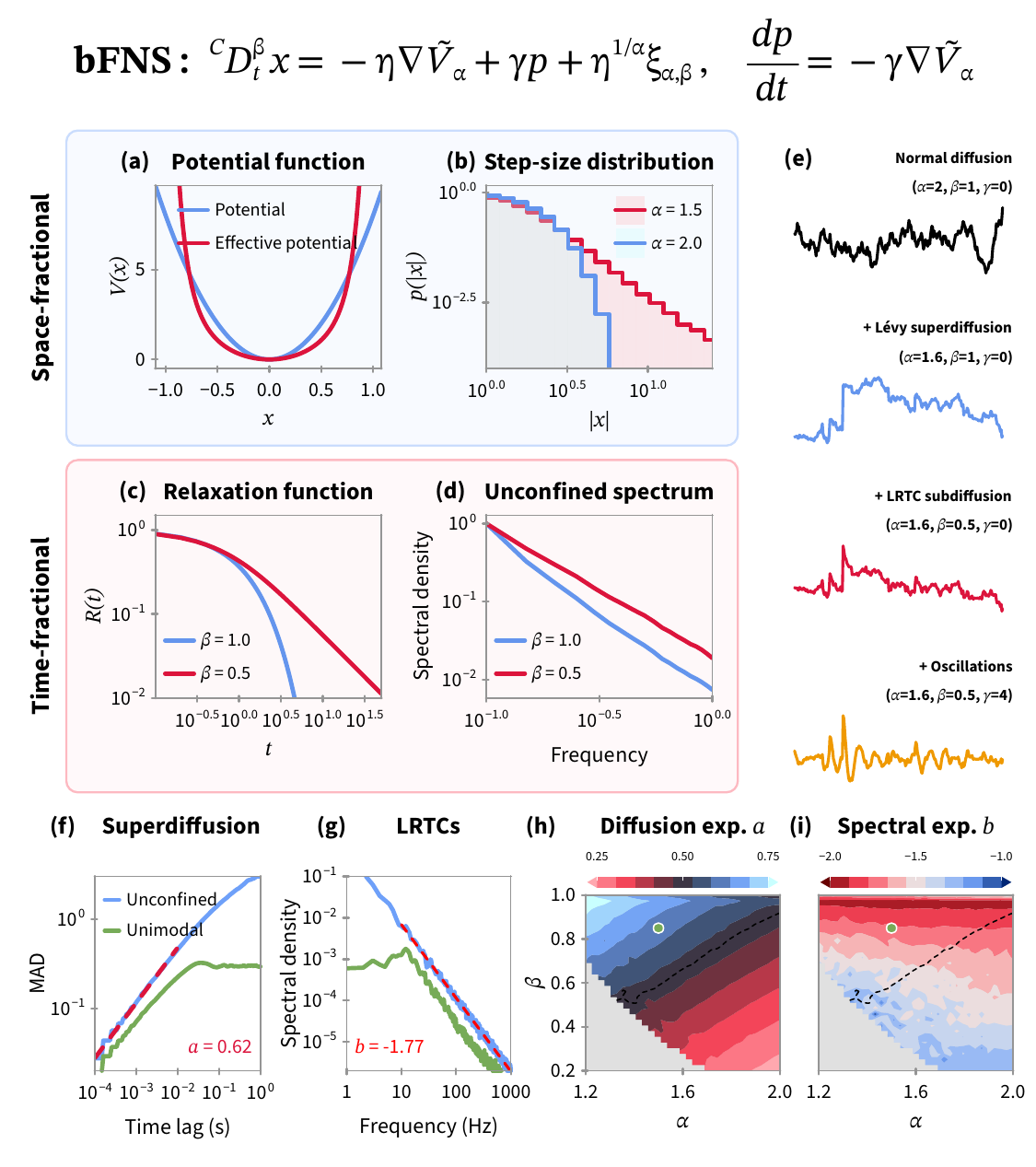}
	{
		\phantomsubcaption\label{subfig:space_fractional_drift}
		\phantomsubcaption\label{subfig:space_fractional_diffusion}
		\phantomsubcaption\label{subfig:time_fractional_drift}
		\phantomsubcaption\label{subfig:time_fractional_diffusion}
		\phantomsubcaption\label{subfig:bFNS_traces}
		\phantomsubcaption\label{subfig:diffusion_exponent}
		\phantomsubcaption\label{subfig:spectral_exponent}
		\phantomsubcaption\label{subfig:diffusion_exponent_sweep}
		\phantomsubcaption\label{subfig:spectral_exponent_sweep}
	}
	\vspace{-1em}\caption{A minimal effective theory links superdiffusion, memory, and oscillations.
	Upper text shows the bFNS model, \cref{eq:bifractional_neural_sampling}.
	\newsubref{subfig:space_fractional_drift} The fractional spatial derivative deforms the potential function (blue), giving an effective potential (red) with a flatter interior and steeper exterior.
	\newsubref{subfig:space_fractional_diffusion} Step-size distributions of the driving noise on log-log axes: heavy, power-law tails for spatial fractional order $\alpha = 1.5$ (red), against the Gaussian case $\alpha = 2$ (blue). \\[0.5em](continued)
	}
	\end{figure*}
	\begin{figure*}
	\ContinuedFloat
	\centering
	\caption{
	\newsubref{subfig:time_fractional_drift} The fractional temporal derivative turns exponential relaxation ($\beta = 1$; blue) into power-law relaxation ($\beta = 0.5$; red) via a power-law memory kernel.
	\newsubref{subfig:time_fractional_diffusion} The fractional temporal derivative correlates the driving fluctuations in time, lifting the unconfined power spectrum above the $f^{-2}$ scaling of normal diffusion ($\beta = 1$; blue) toward shallower slopes ($\beta = 0.5$; red).
	\newsubref{subfig:bFNS_traces} Sample time series showing the influence of each bFNS component: normal diffusion, L\'evy superdiffusion ($\alpha < 2$), long-range temporal correlations (LRTCs; $\beta < 1$), and oscillations ($\gamma > 0$), with parameters annotated.
	\newsubref{subfig:diffusion_exponent} Mean absolute deviation (MAD) of the bFNS model with $\alpha = 1.5$, $\beta = 0.85$, $\gamma = 0.03$, and $\eta = 0.01$, on a flat, unconfined potential (blue) and a unimodal potential (green).
	The unconfined process is superdiffusive across time lags, with its fitted diffusion exponent $a$ annotated (dashed red); the confined process is superdiffusive at short lags before flattening under confinement.
	\newsubref{subfig:spectral_exponent} Power spectral density (PSD) of the same processes, with the fitted spectral exponent $b$ of the unconfined process annotated (dashed red); the confined process shows a peaked spectrum with a power-law high-frequency tail, reflecting local oscillations.
	\newsubref{subfig:diffusion_exponent_sweep} Diffusion exponent $a$ of the unconfined bFNS model across spatial fractional order $\alpha$ and temporal fractional order $\beta$.
	Gray marks invalid parameter combinations, for which the self-similarity exponent lies outside $(0, 1)$; the black dashed line marks the $a = 0.5$ boundary, and the green dot the parameter combination used in \subref{subfig:diffusion_exponent} and \subref{subfig:spectral_exponent}.
	\newsubref{subfig:spectral_exponent_sweep} Spectral exponent $b$ across $\alpha$ and $\beta$, with the same markings.
	}
	\label{fig:theoretical_model}
\end{figure*}
\clearpage
\begin{figure*}[htbp]
	\centering
	\includegraphics[width=\textwidth]{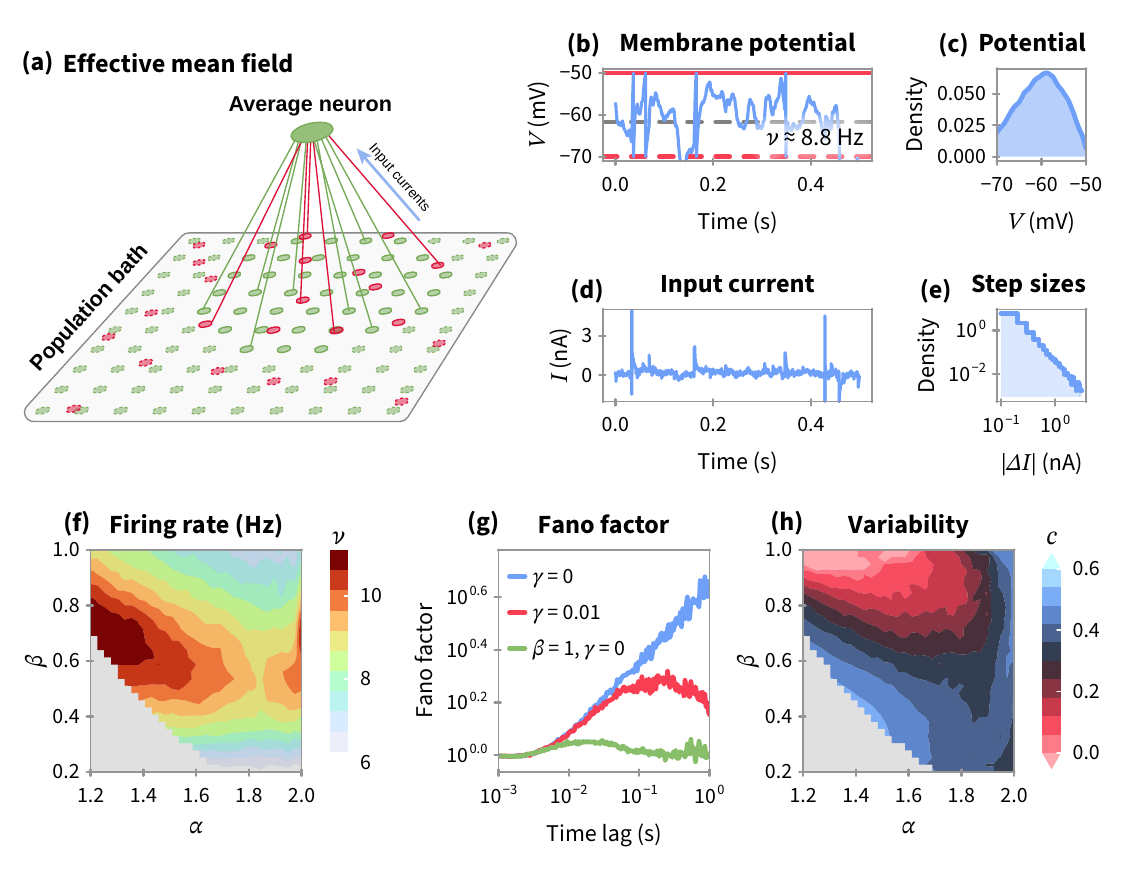}
	{
		\phantomsubcaption\label{subfig:mean_field_schematic}
		\phantomsubcaption\label{subfig:membrane_potential_mean_field}
		\phantomsubcaption\label{subfig:potential_distribution_mean_field}
		\phantomsubcaption\label{subfig:input_current_mean_field}
		\phantomsubcaption\label{subfig:input_distribution_mean_field}
		\phantomsubcaption\label{subfig:firing_rate_mean_field}
		\phantomsubcaption\label{subfig:fano_factor_mean_field}
		\phantomsubcaption\label{subfig:variability_exponent_mean_field}
		}
	\caption{An empirical bi-fractional mean field reproduces the adaptive fractional state.
	\newsubref{subfig:mean_field_schematic} Schematic of our mean-field construction, where a given neuron receives inputs from the surrounding network that can be approximated by the bFNS model.
	\newsubref{subfig:membrane_potential_mean_field} Example membrane potential trace for a simulated mean-field neuron driven by the bFNS model ($\alpha = 1.5$, $\beta = 0.8$, $\eta = 0.03$, $\gamma = 0.01$) with a target density fit from the circuit model. The threshold and reset potentials are marked by solid and dashed red horizontal lines, respectively. The mean potential is marked as a dashed gray line, and the median firing rate $\nu$ is annotated (cf.\ \cref{subfig:membrane_potential,subfig:potential_distribution,subfig:input_distribution} for the circuit model).
	\newsubref{subfig:potential_distribution_mean_field} The distribution of the membrane potential process shown in \subref{subfig:membrane_potential_mean_field}, showing a mean significantly below the spiking threshold.
	\newsubref{subfig:input_current_mean_field} Example mean-field input current trace, generated here by direct simulation of the bFNS model.
	\newsubref{subfig:input_distribution_mean_field} Step-size distribution for the mean-field input current process, shown on log-log axes to highlight power-law tails.
	\newsubref{subfig:firing_rate_mean_field} Firing rate of the mean-field neuron across spatial fractional order $\alpha$ and temporal fractional order $\beta$ (median across $10$ repeated simulations).
	Gray marks invalid parameter combinations, for which the self-similarity exponent lies outside $(0, 1)$.
	\newsubref{subfig:fano_factor_mean_field} Fano factor curves for mean-field neurons in three different regimes about the default parameters used for \subref{subfig:membrane_potential_mean_field}: i) no momentum ($\gamma = 0$; blue); ii) moderate momentum ($\gamma = 0.01$; red); and iii) no long-range memory or momentum ($\beta = 1.0$, $\gamma = 0$; green).
	\newsubref{subfig:variability_exponent_mean_field} Variability exponent $c$ (the exponent of the scaling regime of the Fano factor curve; see \nameref{sec:methods}) across $\alpha$ and $\beta$ for the default parameters, with the same gray masking as \subref{subfig:firing_rate_mean_field}.
	The color scale is clipped to $[0, 0.6]$; values fall below zero near $\beta = 1$.
	}
	\label{fig:mean_field}
\end{figure*}
\clearpage
\begin{figure*}[htbp]
	\centering
	\includegraphics[width=\textwidth]{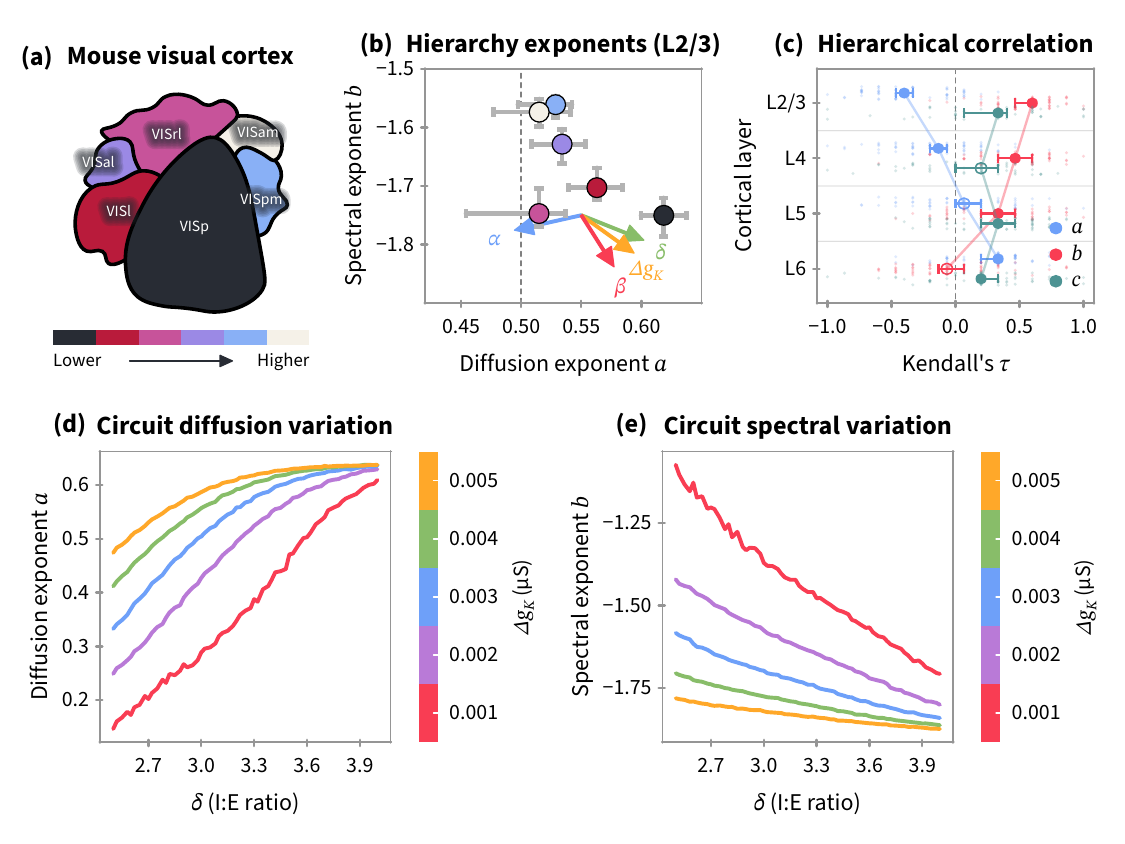}
	{
		\phantomsubcaption\label{subfig:cortex_map}
		\phantomsubcaption\label{subfig:hierarchy_l23}
		\phantomsubcaption\label{subfig:hierarchy_correlation}
		\phantomsubcaption\label{subfig:circuit_diffusion_map}
		\phantomsubcaption\label{subfig:circuit_spectral_map}
	}
	\caption{Working regimes vary systematically across the visual hierarchy.
	\newsubref{subfig:cortex_map} Schematic map of the six visual cortical areas (\texttt{VISp}, \texttt{VISl}, \texttt{VISrl}, \texttt{VISal}, \texttt{VISpm}, \texttt{VISam}), colored by their anatomical hierarchy score~\cite{Harris2019d,Siegle2021} from lower (dark) to higher (light) areas.
	\newsubref{subfig:hierarchy_l23} Diffusion exponent $a$ and spectral exponent $b$ plotted across the six visual areas in layer 2/3 during spontaneous activity, with each area colored as in \subref{subfig:cortex_map}. Points give the across-session bootstrap median, with bars showing the 95\% confidence interval. Arrows indicate the direction in which each model parameter displaces the exponents, as that parameter is increased: the bFNS spatial fractional order $\alpha$ and temporal fractional order $\beta$, and the circuit-model synaptic I:E ratio $\delta$ and excitatory adaptation strength $\Delta g_K$, each labeled in the panel. Arrow lengths are normalized.
	\newsubref{subfig:hierarchy_correlation} Correlation between the dynamical exponents and the cortical hierarchy, by cortical layer. Markers show the across-session median Kendall's $\tau$ for the diffusion exponent $a$ (blue), the spectral exponent $b$ (red), and the variability exponent $c$ (teal), with bars showing 95\% percentile bootstrap intervals over $10^4$ resamples of sessions and faint points showing the per-session values behind each median. Filled markers indicate a Benjamini--Hochberg-adjusted permutation $p < 0.01$.
	The correlations of $a$ and $b$ are strongest in layer 2/3, where the diffusion exponent decreases and the spectral exponent increases with hierarchy score, while the correlation of $c$ is positive at every layer and significant in all but layer 4.
	\newsubref{subfig:circuit_diffusion_map} Diffusion exponent $a$ of the circuit model plotted against the synaptic inhibition-to-excitation ratio $\delta$, with one curve for each of five excitatory adaptation strengths $\Delta g_K$.
	The diffusion exponent saturates near $0.64$ at high $\delta$ (strong inhibition).
	As $\delta$ is lowered through the AF regime, the diffusion exponent decreases below $0.5$ during the transition to the excitation-dominated regime.
	\newsubref{subfig:circuit_spectral_map} Spectral exponent $b$ plotted against $\delta$ for the same values of $\Delta g_K$.
	}
	\label{fig:hierarchical_variation}
\end{figure*}
\clearpage

\section{Tables}

{%
\renewcommand{\arraystretch}{1.5}%
\begin{table}[htbp]
\centering
\caption{Default circuit-model parameters}
\label{tab:params}
\begin{tabular}{lcr}
\hline
	\textbf{Parameter} & \textbf{Symbol} & \textbf{Value} \\
\hline
\multicolumn{3}{l}{\textbf{Neuron parameters}} \\
Membrane capacitance & $C$ & \SI{0.25}{\nano\farad} \\
Leak conductance (E) & $g_L^E$ & \SI{0.0167}{\micro\siemens} \\
Leak conductance (I) & $g_L^I$ & \SI{0.025}{\micro\siemens} \\
Leak reversal & $V_L$ & \SI{-70}{\milli\volt} \\
Potassium reversal & $V_K$ & \SI{-85}{\milli\volt} \\
Spike threshold & $V_{\mathrm{th}}$ & \SI{-50}{\milli\volt} \\
Reset potential & $V_{\mathrm{rt}}$ & \SI{-70}{\milli\volt} \\
Refractory period & $\tau_{\mathrm{ref}}$ & \SI{4}{\milli\second} \\
Adaptation time constant & $\tau_K$ & \SI{40}{\milli\second} \\
Adaptation strength & $\Delta g_K$ & \SI{0.002}{\micro\siemens} \\
\hline
\multicolumn{3}{l}{\textbf{Synaptic parameters}} \\
Excitatory rise time & $\tau_r^E$ & \SI{1.0}{\milli\second} \\
Inhibitory rise time & $\tau_r^I$ & \SI{2.0}{\milli\second} \\
Excitatory decay time & $\tau_d^E$ & \SI{5.0}{\milli\second} \\
Inhibitory decay time & $\tau_d^I$ & \SI{4.5}{\milli\second} \\
Excitatory reversal & $V_E$ & \SI{0}{\milli\volt} \\
Inhibitory reversal & $V_I$ & \SI{-80}{\milli\volt} \\
Excitatory delay & $d_E$ & \SI{1.5}{\milli\second} \\
Inhibitory delay & $d_I$ & \SI{1.5}{\milli\second} \\
\hline
\end{tabular} \hspace{1em} \begin{tabular}{lcr}
\hline
	\textbf{Parameter} & \textbf{Symbol} & \textbf{Value} \\
\hline
\multicolumn{3}{l}{\textbf{Connectivity}} \\
Excitatory density & $\rho$ & \SI{20000}{\per\square\milli\metre} \\
Domain side length & $L$ & \SI{0.5}{\milli\metre} \\
Population E:I ratio & $\zeta$ & 4 \\
Synaptic I:E ratio & $\delta$ & 3.75 \\
E$\to$E radius & $\sigma_{EE}$ & \SI{0.06}{\milli\metre} \\
E$\to$I radius & $\sigma_{EI}$ & \SI{0.07}{\milli\metre} \\
I$\to$E radius & $\sigma_{IE}$ & \SI{0.14}{\milli\metre} \\
I$\to$I radius & $\sigma_{II}$ & \SI{0.14}{\milli\metre} \\
E$\to$E in-degree & $K_{EE}$ & 260 \\
E$\to$I in-degree & $K_{EI}$ & 340 \\
I$\to$E in-degree & $K_{IE}$ & 225 \\
I$\to$I in-degree & $K_{II}$ & 290 \\
E$\to$E weight & $J_{EE}$ & \SI{0.00105}{\micro\siemens} \\
E$\to$I weight & $J_{EI}$ & \SI{0.00145}{\micro\siemens} \\
&& \\
\hline
\multicolumn{3}{l}{\textbf{External input}} \\
External rate & $\nu_{\mathrm{ext}}$ & \SI{10}{\hertz} \\
External synapses & $n_{\mathrm{ext}}$ & 100 \\
&& \\
\hline
\end{tabular}
\vspace{2em}
\end{table}
}

\clearpage

\section{Methods}
\label{sec:methods}

\subsection{Experimental data}
\label{subsec:experimental_data}

We analyze local field potentials (LFPs) and spike data from the Allen Institute Visual Behavior Neuropixels dataset~\cite{Bennett2025}.
This dataset comprises high-density electrophysiological recordings from Neuropixels probes~\cite{Jun2017} in the mouse visual cortex.
LFPs were sampled at \SI{1250}{\hertz} with channels spaced vertically by \SI{40}{\micro\metre}, providing fine laminar resolution across cortical depth.
Probes were positioned in six visual cortical areas: primary visual cortex (\texttt{VISp}) and five higher visual areas (\texttt{VISl}, \texttt{VISrl}, \texttt{VISal}, \texttt{VISpm}, and \texttt{VISam}).
Recordings were obtained from $81$ mice during passive viewing of \SI{250}{\milli\second} visual stimuli, including gray screens, luminance flashes, and natural images; for our analysis, we use the gray-screen epochs to characterize spontaneous activity.
We filter down to $69$ sessions over $41$ mice by requiring that every session intersect all six target areas, provide LFP on all six probes, and carry annotated target-structure locations, while excluding sessions flagged for abnormal histology or abnormal activity and four sessions whose cortical surface position could not be identified.
For the Fano factor analyses, we retain units that were not classified as noise during spike sorting, were present for more than $90\%$ of the recording, and had an interspike-interval violation score below $0.5$~\cite{Siegle2021}.
We omit the amplitude cutoff of the standard Allen quality filter, since applying it leaves most layer 2/3 areas and sessions short of the five units required for a Fano factor fit.
For each area, we take the anatomical hierarchy score derived from the feedforward or feedback character of interareal projections~\cite{Harris2019d}, as previously applied to these six areas~\cite{Siegle2021}.

\subsection{Spiking neural circuit model}

We model a spatially embedded recurrent spiking neural network consisting of excitatory (E) and inhibitory (I) neuronal populations with conductance-based synapses and spike-frequency adaptation in the excitatory population; the full set of parameter values is summarized in \cref{tab:params}.
All simulations were performed with the \href{https://github.com/brendanjohnharris/Dewdrop.jl}{Dewdrop.jl} Julia package~\cite{Harris2026j} on a GPU backend, using a fixed time step of \SI{0.1}{\milli\second} and a discarded transient period of \SI{5}{\second}; the example simulation of \cref{fig:experimental_data} was run for \SI{55}{\second}, while each realization of the parameter sweeps described below was run for \SI{35}{\second}.

\subsubsection{Single-neuron dynamics}

Both excitatory and inhibitory neurons follow a leaky integrate-and-fire model with spike-frequency adaptation in the excitatory population~\cite{Treves1993}.
The membrane potential $V_i$ of neuron $i$ evolves according to:
\begin{equation}
    C \frac{dV_i}{dt} = -g_L(V_i - V_L) - g_{K,i}(V_i - V_K) + I_{i}\,,
\end{equation}
where $C$ is the membrane capacitance, $g_L$ is the leak conductance (larger for inhibitory than excitatory neurons, reflecting the higher intrinsic excitability of interneurons; \cref{tab:params}), $V_L$ is the leak reversal potential, $g_{K,i}$ is the spike-triggered potassium conductance providing adaptation, $V_K$ is the potassium reversal potential, and $I_{i}$ is the total synaptic input current.

The adaptation conductance $g_{K,i}$ decays exponentially with time constant $\tau_K$ and is incremented by $\Delta g_K$ following each spike:
\begin{equation}
    \frac{dg_{K,i}}{dt} = -\frac{g_{K,i}}{\tau_K} + \Delta g_K \sum_k \delta(t - t_i^{(k)})\,,
\end{equation}
where $t_i^{(k)}$ denotes the $k$-th spike time of neuron $i$.
For inhibitory neurons, $\Delta g_K = 0$, so they exhibit no spike-frequency adaptation.

When the membrane potential reaches the threshold $V_{\mathrm{th}}$, the neuron emits a spike, the membrane potential is reset to $V_{\mathrm{rt}}$, and the neuron enters an absolute refractory period of duration $\tau_{\mathrm{ref}}$ during which the membrane potential is clamped.
Membrane potentials are initialized by drawing from a uniform distribution $V_i(0) \sim \mathcal{U}(V_{\mathrm{rt}}, V_{\mathrm{th}})$.

\subsubsection{Synaptic dynamics}

Synaptic transmission is modeled using conductance-based synapses with dual-exponential kinetics.
For a synapse from presynaptic neuron $j$ to postsynaptic neuron $i$, the synaptic conductance $g_{ij}(t)$ evolves according to a pair of coupled differential equations:
\begin{align}
    \frac{dg_{ij}}{dt} &= -\frac{g_{ij}}{\tau_d} + h_{ij}, \\
    \frac{dh_{ij}}{dt} &= -\frac{h_{ij}}{\tau_r} + A \left(\frac{1}{\tau_r} - \frac{1}{\tau_d}\right) \bar{g}_{ij} \sum_k \delta(t - t_j^{(k)} - d_{ij})\,,
\end{align}
where $h_{ij}$ is an auxiliary variable governing the rise phase, $\bar{g}_{ij}$ is the maximal synaptic conductance, $t_j^{(k)}$ is the $k$-th spike time of the presynaptic neuron, $d_{ij}$ is the synaptic transmission delay, $\tau_r$ is the rise time constant, and $\tau_d$ is the decay time constant.
The normalization factor
\begin{equation}
    A = \frac{\tau_d}{\tau_d - \tau_r} \left(\frac{\tau_r}{\tau_d}\right)^{\tau_r/(\tau_r - \tau_d)}
\end{equation}
ensures that the peak conductance following a single spike equals $\bar{g}_{ij}$.

The total synaptic current to neuron $i$ is given by:
\begin{equation}
    I_{i} = -\sum_{j \in E} g_{ij}^E(t)(V_i - V_E) - \sum_{j \in I} g_{ij}^I(t)(V_i - V_I)\,,
\end{equation}
where the sums run over all excitatory (E) and inhibitory (I) presynaptic partners, and $V_E$ and $V_I$ are the reversal potentials for excitatory and inhibitory synapses, respectively.

\subsubsection{Connectivity}

Neurons are distributed on a two-dimensional square domain of side length $L$ with periodic boundary conditions.
Excitatory neurons are positioned on a regular grid with density $\rho$, yielding $N_E = \rho L^2$ excitatory neurons arranged on an $n_E \times n_E$ grid where $n_E = L\sqrt{\rho}$.
Inhibitory neurons are uniformly distributed at random positions with density $\rho/\zeta$, giving $N_I = N_E/\zeta$ inhibitory neurons, where $\zeta$ is the population excitatory-to-inhibitory (E:I) ratio; we use $\zeta$ here rather than $\gamma$, which denotes the momentum coupling of the bi-fractional model.

Synaptic connections are established probabilistically based on the Euclidean distance between neurons with periodic boundaries.
The connection probability from neuron $j$ at position $\symbf{x}_j$ to neuron $i$ at position $\symbf{x}_i$ follows an exponential kernel:
\begin{equation}
    p_{ij} = p_{\max}^{XY} \exp\left(-\frac{r_{\,ij}}{\sigma_{XY}}\right)\,,
\end{equation}
where $r_{\,ij} = \|\symbf{x}_i - \symbf{x}_j\|_{\mathbb{T}^2}$ is the toroidal distance between neurons, $X,Y \in \{E, I\}$ denote the pre- and postsynaptic population types respectively, $\sigma_{XY}$ is the spatial scale of connectivity, and $p_{\max}^{XY}$ is the maximum connection probability.

Rather than specifying $p_{\max}^{XY}$ directly, we parameterize the connectivity by the mean number of incoming connections $K_{XY}$ per postsynaptic neuron.
The total mass of the probability kernel over the domain is approximated by disregarding periodic boundaries as:
\begin{equation}
    \omega_{XY} = 2\pi \sigma_{XY}^2 p_{\max}^{XY}\,,
\end{equation}
from which the expected in-degree is $\langle k_{XY} \rangle = \rho_X \omega_{XY}$, where $\rho_X$ is the density of the presynaptic population.
Setting $\langle k_{XY} \rangle = K_{XY}$ determines:
\begin{equation}
    p_{\max}^{XY} = \frac{K_{XY}}{2\pi \sigma_{XY}^2 \rho_X}\,.
\end{equation}
The total number of synaptic connections in projection $X \to Y$ is then $K_{XY} N_Y$, and these are sampled using weighted Gumbel-max sampling to ensure the exact number of connections while respecting the distance-dependent probabilities~\cite{Kool2019a}.
Fixing the total number of connections substantially reduces the cost of constructing the connectome on the GPU by allowing arrays to be sized statically.

\subsubsection{Synaptic weights}

Synaptic weights are scaled to ensure that recurrent dynamics remain in a balanced regime~\cite{Ni2024}.
The mean weight of synapses onto postsynaptic neuron $i$ is inversely proportional to the square root of its in-degree $k_i$:
\begin{equation}
    \langle w_{ij} \rangle_j = \frac{J^{XY}_{\mathrm{rec}}}{\sqrt{k_i}}\,,
\end{equation}
where the constant $J^{XY}_{\mathrm{rec}}$ is computed to ensure that the population-averaged total synaptic weight equals the target weight $J_{XY}$:
\begin{equation}
    J^{XY}_{\mathrm{rec}} = J_{XY} \frac{\sum_i k_i}{\sum_i \sqrt{k_i}}\,.
\end{equation}
Individual synaptic weights are then drawn from a normal distribution $w_{ij} \sim \mathcal{N}(\langle w_{ij} \rangle, 0.2\langle w_{ij} \rangle)$ to introduce heterogeneity while preserving the mean.

The baseline synaptic weights for excitatory projections are $J_{EE}$ and $J_{EI}$.
Inhibitory synaptic weights are parameterized through a synaptic I:E ratio $\delta$, scaled by the ratio of the corresponding in-degrees, giving
\begin{equation}
    J_{IE} = \delta \frac{K_{EE}}{K_{IE}} J_{EE}, \quad J_{II} = \delta \frac{K_{EI}}{K_{II}} J_{EI}\,.
\end{equation}
The in-degree factors make $\delta$ the ratio of the total recurrent inhibitory to excitatory drive onto each population, rather than the ratio of individual synaptic weights; this parameterization allows systematic exploration of the excitation--inhibition balance by varying a single parameter, which we fix at $\delta = 3.75$ except where the sweeps of \nameref{subsec:circuit_sweeps} vary it.

\subsubsection{External input}

Each neuron receives external excitatory input from a population of $N_{\mathrm{ext}} = \sqrt{n_{\mathrm{ext}} N_E}$ Poisson neurons firing at rate $\nu_{\mathrm{ext}}$, each connected to neurons in both populations with probability $p_{\mathrm{ext}} = \sqrt{n_{\mathrm{ext}}/N_E}$, such that each neuron receives $n_{\mathrm{ext}}$ external synapses on average.
The excitatory and inhibitory populations share a common-mode drive: the external neurons emit a single set of spike trains that is delivered to both populations, with connectivity and weights drawn independently for each.
Recurrent inhibition tracks and cancels this common mode, suppressing population-wide fluctuations relative to an independent drive.
External synapses use the same dual-exponential kinetics as recurrent excitatory synapses, with synaptic weights $J_{EE}$ for external-to-excitatory connections and $J_{EI}$ for external-to-inhibitory connections.

\subsection{Circuit parameter sweeps}
\label{subsec:circuit_sweeps}

To characterize how the dynamical exponents depend on circuit parameters, we sweep the spiking circuit model over two-dimensional planes passing through its default working point (\cref{tab:params}), varying two parameters at a time and holding all others fixed.
For the plane shown in \cref{fig:hierarchical_variation}, we vary the synaptic I:E ratio $\delta$ over $61$ values spanning $[2.5, 4]$ and the excitatory adaptation strength $\Delta g_K$ over $21$ values spanning \SIrange{0.001}{0.005}{\micro\siemens}.
At each grid point we simulate $10$ independent realizations of the network, differing in the random connectome and initial conditions, each for \SI{35}{\second} with a time step of \SI{0.1}{\milli\second} and a discarded transient of \SI{5}{\second}.
For every tenth excitatory neuron we then compute the MAD curve of the synaptic input current, over $\sim$100 logarithmically spaced lags spanning \SIrange{1}{1000}{\milli\second}, along with its power spectral density.
We extract the diffusion and spectral exponents of each sampled neuron with the estimators described in \nameref{subsec:fitting_exponents}: a single-component MAPPLE fit to the MAD curve over lags up to \SI{8}{\milli\second}, and a one-component, one-peak MAPPLE fit to the PSD over \SIrange{10}{1000}{\hertz}.
Each point of \cref{subfig:circuit_diffusion_map,subfig:circuit_spectral_map} shows the median exponent pooled across neurons and realizations; grid cells with fewer than five completed realizations were discarded.
The 95\% confidence intervals quoted at the working point are computed across network realizations: each realization is reduced to the median exponent over the neurons sampled from it, and the interval spans the 2.5th and 97.5th percentiles of $10^4$ bootstrap resamples of these (at most $10$) per-realization medians.
Because realizations differ in connectome and initial conditions, this interval captures run-to-run variability of the network median; it is distinct from the across-neuron intervals quoted for the single long simulation of \cref{fig:experimental_data}, which bootstrap the median over all fitted neurons within one realization and so measure the precision of that population median.

The direction arrows in \cref{subfig:hierarchy_l23} summarize the local effect of each parameter on the pair of exponents.
For each circuit parameter, we compute the net displacement in the $(a, b)$ plane between the ends of a local interval ($\delta: 2.95 \to 3.45$, centered on the region that overlaps the experimentally observed exponents; $\Delta g_K: 0.001 \to \SI{0.003}{\micro\siemens}$, centered on the default working point), averaged over the other parameter restricted to its own interval; for the bFNS model, we compute the same displacement over $\alpha: 1.35 \to 1.65$ and $\beta: 0.75 \to 0.95$ using the unconfined sweep shown in \cref{subfig:diffusion_exponent_sweep,subfig:spectral_exponent_sweep}.
Each arrow is then normalized to a fixed length in display coordinates.
The exponents plotted in \cref{subfig:hierarchy_l23} were estimated with the MAPPLE estimators detailed in \nameref{subsec:fitting_exponents}: a single-component fit to the MAD curve over the short-lag band and, for the experimental PSD, two power-law components and two Gaussian peaks over \SIrange{1}{500}{\hertz} with the interior transition held at \SI{3}{\hertz}.
Points show the across-session median, with 95\% confidence intervals from $10^4$ bootstrap resamples.

\subsection{Theoretical model}
\label{sec:theoretical_model_sup}

We propose a fully bi-fractional stochastic process with both spatial and temporal fractional derivatives.
Our full theoretical model is described by the following Langevin-like equation on the one-dimensional position $x$ and auxiliary momentum $p$:
\begin{equation}
	\label{eq:bifractional_neural_sampling_sup}
	\begin{aligned}
		{}^C D_t^\beta x &= -\eta \nabla \tilde{V}_\alpha + \gamma p + \eta^{\frac{1}{\alpha}} \xi_{\alpha,\beta}\,,\\
		D_t p &= -\gamma \nabla \tilde{V}_\alpha \,,
	\end{aligned}
\end{equation}
where $\alpha \in (1, 2]$ is the spatial fractional order, $\beta \in (0, 1]$ is the temporal fractional order, $\eta$ is the noise intensity controlling the timescale of diffusion, $\gamma$ is the momentum coupling parameter, and $\nabla$ is the regular spatial gradient operator.
The operator $D_t$ represents the standard temporal derivative, whereas ${}^C D_t^\beta$ is the Caputo fractional temporal derivative of order $\beta$, defined as:
\begin{equation}
	\label{eq:caputo_derivative}
    {}^C D_t^\beta[f(t)] = \frac{1}{\Gamma(1-\beta)} \int_0^t \frac{f'(\tau)}{(t-\tau)^\beta} d\tau\,,
\end{equation}
where $\Gamma(z) = \int_0^\infty \exp(-u) u^{z-1} du\,, \Re{z} > 0\,,$ is the gamma function.
Here $\xi_{\alpha,\beta}(t)$ is linear fractional stable noise, the increment process of a linear fractional stable motion $X(t)$~\cite{Watkins2009,Mazur2020}, centered at zero with unit scale (giving a variance of $2$ for $\alpha = 2$, $\beta = 1$), a L\'evy order of $\alpha$, and a self-similarity relationship $X(st) \overset{d}{=} s^H X(t)$ (where $\overset{d}{=}$ denotes distributional equality).
We make the heuristic choice to use a self-similarity exponent $H = 1/2 + 1/\alpha - \beta/2$, which empirically produces a correct stationary distribution; future work should aim to verify this choice of $H$ analytically by deriving a full bi-fractional Fokker--Planck equation for \cref{eq:bifractional_neural_sampling}.
This self-similarity exponent must lie in $(0, 1)$, and requires the noise amplitude to scale as $\Delta t^{H}$ in the numerical discretization.

The function $\tilde{V}_\alpha(x)$ is the effective potential (illustrated in \cref{subfig:space_fractional_drift} for a unimodal potential) that captures the influence of the fractional spatial derivative on the deterministic component of the system.
The effective potential is defined through its gradient as:
\begin{equation}
	\label{eq:effective_potential_gradient_sup}
    \nabla \tilde{V}_\alpha(x) = \frac{\nabla [\mathcal{D}_x^{\alpha - 2}\pi]}{\pi}\,,
\end{equation}
where $\mathcal{D}_x^{\alpha - 2}$ is the Riesz fractional derivative of order $\alpha - 2$, given in the Fourier domain by:
\begin{equation}
	\label{eq:riesz_derivative}
    \widehat{\mathcal{D}_x^{\alpha}f}(k) = -|k|^\alpha \hat{f}(k)\,.
\end{equation}
The effective potential reduces to the classical Langevin potential---$V(x) = -\log \pi(x)$, where $\pi(x)$ is the target distribution to be sampled---when $\alpha = 2$.
Crucially, our formulation maintains the stationary distribution $\pi(x)$, as shown numerically in \cref{subfig:stationary_distribution_unimodal,subfig:stationary_distribution_bimodal}; we verify that the sampling error remains low across a range of fractional orders $\alpha$ and $\beta$ in \cref{subfig:sampling_accuracy_unimodal} (unimodal potential) and \cref{subfig:sampling_accuracy_bimodal} (bimodal potential).

\subsection{bFNS implementation}
\label{sec:bfns_implementation}

Here we describe the numerical schemes used to simulate the full bi-fractional FNS model; our implementation is available in the Julia package \href{https://github.com/brendanjohnharris/FractionalNeuralSampling.jl}{FractionalNeuralSampling.jl}~\cite{Harris2026i}.
All simulations of the theoretical model were performed using an Euler--Maruyama L1 method (detailed below) with a fixed time step of \SI{0.1}{\milli\second}, a duration of \SI{25}{\second}, and a discarded transient period of \SI{5}{\second}.

\subsubsection{Driving noise}

We use linear fractional stable noise~\cite{Stoev2004} as the stochastic drive for the theoretical model, which is a stationary process with heavy-tailed (stable) marginal distributions and long-range dependence characterized by a power-law power spectral density.
Linear fractional stable noise is defined by the increments of a linear fractional stable motion $X_t$~\cite{Watkins2009,Mazur2020}, given by:
\begin{equation}
X_t = \int_{\mathbb{R}} \left\{ (t-s)^{H-1/\alpha}_+ - (-s)^{H-1/\alpha}_+ \right\} dL_s\,, \quad x_+ := \max\{x, 0\}\,,
\end{equation}
where $L_s$ is a symmetric $\alpha$-stable L\'evy motion with scaling parameter $\sigma > 0$ and self-similarity exponent $H \in (0,1)$.

The fractional orders of the distribution and temporal correlations of the driving noise process must match the orders of the spatial and temporal fractional derivatives in the model, such that the empirical stationary distribution matches the target Gibbs distribution.
Specifically, noise distributed according to an $\alpha$-stable distribution corresponds to a fractional spatial order of $\alpha$, balancing the fractional spatial derivative~\cite{Ye2018}.
To balance the fractional temporal derivative~\cite{Fang2020}, we choose the self-similarity exponent of the integrated noise process as $H = 1/2 + 1/\alpha - \beta/2$, where $\beta$ is the temporal fractional order.
This heuristic self-similarity exponent for the integrated noise process is the simplest additive combination of two established requirements for satisfying generalized fluctuation--dissipation: i) that $H_\beta = 1 - \beta/2$ for time-fractional Langevin equations~\cite{Li2017b}; and ii) that $H_\alpha = 1/\alpha$ for space-fractional processes~\cite{Dybiec2012}.

We simulate linear fractional stable noise using the FFT-based approach detailed in \citet{Mazur2020}, choosing a subsampling factor of $m=128$ and a lookback window length of \SI{1}{\second} at a timestep of \SI{0.1}{\milli\second}.
As a final step, we normalize the generated noise to have a median of $0$ and an interquartile range consistent with a Gaussian distribution of variance $2$.

\subsubsection{Fractional spatial derivative}

We use a spectral method to approximate the drift term $\nabla \tilde{V}_\alpha = \nabla [\mathcal{D}_x^{\alpha - 2}\pi]/\pi$ on a given periodic domain.
First, we represent the target distribution $\pi\approx\hat{\pi}$, regular derivative $\nabla\approx\hat{\nabla}$, and the negative Laplacian $-\Delta \approx -\hat{\Delta}$ in the Fourier domain using $N \approx 1000$ modes~\cite{Olver2014}, where $\hat{\cdot}$ indicates the spectral approximation of the corresponding function or operator.
In the Fourier domain, the Laplacian operator is diagonal and negative definite with eigenvalues $\widehat{\Delta}(k) = -|k|^2$.
By taking the negative Laplacian and raising each diagonal entry to the power $(\alpha - 2)/2$, we obtain the approximate fractional Laplacian operator $(-\hat{\Delta})^{\frac{\alpha - 2}{2}}$.
For $\alpha < 2$ this exponent is negative, so the zero-frequency entry is singular; we set it to zero rather than raising zero to a negative power.
The zero mode contributes only a constant offset to $\mathcal{D}_x^{\alpha - 2}\pi$, which is removed by the subsequent gradient, so we apply the fractional operator before differentiating.
To avoid numerical instability when $\pi$ is close to zero, we add a small regularization constant $\lambda = 10^{-4}$ to the denominator of the effective-potential gradient, \cref{eq:effective_potential_gradient_sup}.
Given $\mathcal{D}_x^{\alpha - 2} = -(-\Delta)^{(\alpha - 2)/2}$ in one dimension, our numerical implementation of \cref{eq:effective_potential_gradient_sup} then reads:
\begin{equation}
	\nabla \tilde{V}_\alpha(x) \approx \frac{-\hat{\nabla} [(-\hat{\Delta})^{\frac{\alpha - 2}{2}}\hat{\pi}](x)}{\hat{\pi}(x) + \lambda}\,.
\end{equation}


\subsubsection{Fractional temporal derivative}

To approximate the fractional temporal derivative ${}^C D_t^\beta$ in \cref{eq:bifractional_neural_sampling_sup}, we use the L1 finite-difference scheme~\cite{Lin2007}.

First, we consider the fractional stochastic differential equation:
\begin{equation}
    {}^C D_t^\beta x(t) = f(x(t)) + g\xi(t)\,, \quad t > 0\,,
\end{equation}
where ${}^C D_t^\beta$ denotes the Caputo fractional derivative of order $\beta \in (0, 1]$ [see \cref{eq:caputo_derivative}], $f: \mathbb{R}^d \to \mathbb{R}^d$ is the drift coefficient, $g$ is the diffusion coefficient (the constant $\eta^{1/\alpha}$ for bFNS), $\xi(t)$ is a noise process (typically white noise), and $x(t) \in \mathbb{R}^d$ is the state variable.
When $\beta = 1$, this reduces to the standard It\^o stochastic differential equation, and the L1 scheme reduces to the classical Euler--Maruyama method.

The L1 scheme approximates the Caputo derivative using a weighted sum of historical differences.
For $j = 1, 2, \ldots, n$, the L1 weights are:
\begin{equation}
    w_j = (j+1)^{1-\beta} - j^{1-\beta}\,.
\end{equation}
Note that $w_0 = 1$, which is separated out and incorporated into a correction factor for the Euler--Maruyama update.
This correction factor, given by:
\begin{equation}
    \chi = \Gamma(2-\beta) (\Delta t)^{\beta - 1}\,,
\end{equation}
is added to ensure proper scaling of the drift and diffusion terms in the fractional case; here, $\Delta t$ is the time step.

Given the current state $x_n$ at time $t_n$, the next state $x_{n+1}$ is computed as:
\begin{equation}
\begin{aligned}
    x_{n+1} &= x_n + \left[ \chi \, f(x_n) \Delta t\right] + \left[ \chi \, g \, \Delta L_n \right] - \left[ \sum_{j=1}^{n} w_j \Delta x_{n-j}\right]\,,
\end{aligned}
\end{equation}
where the $w_j$ are the L1 weights (indexed in reverse order, so that $j=1$ corresponds to the most recent step), $\Delta x_{n-j} = x_{n-j+1} - x_{n-j}$ are the historical increments, and $\Delta L_n$ is the increment of the driving noise over the step.
For a self-similar drive of exponent $H$ this increment scales as $\Delta t^{H}$, recovering the Brownian $\sqrt{\Delta t}$ only when $H = 1/2$.

This update can be decomposed into three terms: a corrected EM drift $\chi \, f(x_n) \Delta t$, a corrected EM diffusion $\chi \, g \, \Delta L_n$, and a history contribution term $-\sum_{j=1}^{n} w_j \Delta x_{n-j}$.
The history contribution term encodes the long-range memory of the fractional temporal derivative: at each time step, the update depends on all previous state increments, weighted by the slowly decaying L1 coefficients $w_j \sim j^{-\beta}$ for large $j$.
In bFNS only the position $x$ carries a fractional derivative, so the correction factor and the history term are applied to $x$ alone; the momentum $p$ is advanced by a standard Euler--Maruyama step.
For the bFNS model, we use a truncated history length of \SI{1}{\second} to match the window used for linear fractional stable noise.

\subsection{Empirical mean field}
\label{subsec:mean_field_model_sup}

We construct the mean-field approximation by considering a single neuron receiving statistically typical inputs from the surrounding `bath' of neurons (illustrated in \cref{subfig:mean_field_schematic}).
The dynamics of the mean-field neuron follow similar equations as the circuit-model neurons:
\begin{equation}
	\begin{aligned}
		\label{eq:mean_field_neuron}
		C \frac{dV}{dt} &= -g_L(V - V_L) - g_K(V - V_{K}) + I(t)\,,\\
		\frac{dg_K}{dt} &= -\frac{g_K}{\tau_K} + \Delta g_K \delta(t - t_s) \,,
	\end{aligned}
\end{equation}
where $V$ is the membrane potential (\unit{\milli\volt}), $g_K$ is the potassium conductance that implements spike-triggered adaptation (\unit{\micro\siemens}), and $I(t)$ is the total synaptic input (summing background and recurrent inputs; \unit{\nano\ampere}).
The constant parameters of the membrane potential are the membrane capacitance $C = \SI{0.25}{\nano\farad}$, the leak conductance $g_L = \SI{0.0167}{\micro\siemens}$, the leak reversal potential $V_L = \SI{-70}{\milli\volt}$, and the potassium reversal potential $V_K = \SI{-85}{\milli\volt}$.
The adaptation variable $g_K$ undergoes spike-triggered increase and exponential decay, where $\delta(t - t_s)$ is the Dirac delta function centered on each spike time $t_s$ (so that $g_K$ jumps by $\Delta g_K$ at each spike), $\Delta g_K = \SI{0.002}{\micro \siemens}$ is the spike-triggered adaptation increase, and $\tau_K = \SI{40}{\milli\second}$ is the adaptation time constant.
Spikes are emitted when the membrane potential crosses the threshold $V_\textrm{th} = \SI{-50}{\milli\volt}$, at which point the membrane potential is reset to $V_\textrm{rt} = \SI{-70}{\milli\volt}$.
Crucially, all terms in \cref{eq:mean_field_neuron} except for the input $I(t)$ are deterministic; we simulate the model at a timestep of $dt = \SI{0.1}{\milli\second}$, matching the circuit model (the example neuron of \cref{subfig:membrane_potential_mean_field} uses a refined solver step of \SI{0.05}{\milli\second}, saved at \SI{0.1}{\milli\second}).
We use bFNS as an approximation for this net synaptic input $I(t)$, giving an empirical mean-field model.
We simulate this model using the same numerical schemes detailed in \cref{sec:bfns_implementation}; the example neuron of \cref{subfig:membrane_potential_mean_field,subfig:potential_distribution_mean_field,subfig:input_current_mean_field,subfig:input_distribution_mean_field} was run for an extended duration of \SI{55}{\second}, whereas the parameter sweeps behind \cref{subfig:firing_rate_mean_field,subfig:fano_factor_mean_field,subfig:variability_exponent_mean_field} retain the \SI{25}{\second} duration given above.

\subsection{Fitting dynamical exponents}
\label{subsec:fitting_exponents}

To extract dynamical exponents from time-series data, we fit a Multiple Adaptive Peaks and Power-Law Exponents (MAPPLE) model to the empirical curves of interest.
The MAPPLE model decomposes a curve into a sum of piecewise power-law segments with smooth transitions, plus Gaussian peaks, characterizing the scale-free and oscillatory components of neural activity within a single fit.
We fit MAPPLE models to all MAD and PSD curves, across the experimental data, circuit model, and bFNS model, to estimate the diffusion and spectral exponents; the variability exponent is estimated from the Fano factor as detailed below.
We use MAPPLE as a generalization of the field-standard FOOOF model~\cite{Donoghue2020}; FOOOF only fits a single power-law component with a low-frequency knee, whereas MAPPLE fits an arbitrary number of components, as required for MAD and Fano factor curves.
Our implementation is available in the Julia package \href{https://github.com/brendanjohnharris/TimeseriesTools.jl}{TimeseriesTools.jl}~\cite{Harris2026h}.

Before fitting, we resample each power spectrum onto a logarithmically spaced frequency grid, taking the geometric mean within equal-width log-frequency bins; this equalizes the sample density across decades and averages down the high-frequency estimation noise that would otherwise dominate the fit.
For MAD curves, which are not resampled, we instead weight the squared residuals by the log-spacing of the samples, so that the objective approximates an integral over log-lag rather than a sum over samples.

\subsubsection{MAPPLE model specification}

For a given frequency $f$ (or time lag $\tau$), the MAPPLE model predicts a spectral density (or mean absolute deviation) $\hat{S}(f)$ as:
\begin{equation}
    \hat{S}(f) = \sum_{i=1}^{N_c} w_i(f) \cdot A_i \cdot f^{b_i} + \sum_{j=1}^{N_p} P_j \exp\left[-\frac{(f - f_j)^2}{2\sigma_j^2}\right]\,,
\end{equation}
where $N_c$ is the number of power-law components, $A_i$ is the amplitude of the $i$-th component, $b_i$ is the power-law exponent of the $i$-th component, $w_i(f)$ is a smooth windowing function that determines the frequency range over which each component is active, and $N_p$ is the number of Gaussian peaks, each with height $P_j$, center frequency $f_j$, and width $\sigma_j$.
Peak widths are parameterized as $\sigma_j = f_j \tanh(s_j)$, where $s_j$ is a fitted width in log-frequency space, so that each peak maintains an approximately constant width in log-frequency.

The windowing function $w_i(f)$ is defined to provide smooth transitions between adjacent power-law components:
\begin{equation}
    w_i(f) = \frac{1}{4}\left[1 + \tanh\left(\frac{\log_{10}f - \log_{10}f_i^{(1)}}{\varepsilon}\right)\right] \cdot \left[1 + \tanh\left(\frac{\log_{10}f_i^{(2)} - \log_{10}f}{\varepsilon}\right)\right]\,,
\end{equation}
where $f_i^{(1)}$ and $f_i^{(2)}$ are the start and stop frequencies for the $i$-th component, and $\varepsilon$ is the transition width in log-frequency space; when a single power-law component is fit, the windowing is omitted and the aperiodic component reduces to a pure power law.
In multi-component models, the amplitudes of successive components are constrained to maintain continuity at transition points: $A_{i+1} = A_i (f_i^{(2)})^{b_i - b_{i+1}}$.

\subsubsection{Initial parameter estimation}

We initialize the power-law components by performing simple linear regression on the log-log curve to estimate an overall slope, assigning this common slope to every component and distributing the transition points evenly across the fitted log-frequency range.
The transition width $\varepsilon$ is initialized to one-twentieth of the fitted log-frequency span, and the overall amplitude to the value of the lowest-frequency sample.

To initialize the peaks, we first estimate the aperiodic background as a continuous piecewise-linear function of log-frequency, fit robustly by iteratively discarding positive excursions and refitting through the remaining lower envelope.
The requested number of peaks is then seeded from the most prominent local maxima of the residual, after rejecting candidate peaks narrower than the local frequency resolution or wider than half the fitted range; each peak's height is initialized from its prominence above the local background, and its width from the extent of the detected maximum.

\subsubsection{Parameter optimization}

We refine the initial parameters using bounded optimization with the L-BFGS algorithm~\cite{Liu1989}, with box constraints enforced by a log-barrier method and gradients obtained by forward-mode automatic differentiation.
The objective function is the weighted sum of squared residuals in log-space:
\begin{equation}
    \mathcal{L}(\symbfit{\theta}) = \sum_{k=1}^{N_f} \omega_k \left[\log_{10}\hat{S}(f_k; \symbfit{\theta}) - \log_{10}S(f_k)\right]^2\,,
\end{equation}
where $\symbfit{\theta}$ represents all model parameters (component exponents and transition frequencies, peak parameters, overall amplitude, and transition width), $N_f$ is the number of frequency points, $S(f_k)$ is the empirical spectral density at frequency $f_k$, and the weights $\omega_k$ are uniform for log-resampled spectra and proportional to the log-frequency spacing of the samples for MAD curves (as above).
During refinement, each transition frequency is constrained to the fitted range, with its lower bound raised by one-sixth of the fitted log-frequency span so that the leading component retains a minimum width, and the transition width is constrained between one-quarter of the minimum log-frequency spacing and one-third of the fitted span; together these bounds prevent components from collapsing in width.
The initialization is kept as a candidate fit, then each model is refined twice, once with peak centers confined near their detected frequencies and once with peaks free over the whole range, before the solution with the lowest objective is retained.

\subsubsection{Extraction of dynamical exponents}

For the mean absolute deviation and the power spectral density, the number of components and peaks is fixed in advance to match the shape of the curve, and the dynamical exponent of interest is extracted from the power-law component spanning the relevant scale-free regime; the Fano factor fits, which differ between the experimental data and the models, are described separately below.

To measure the diffusion exponent $a$, we fit a single power-law component (no peaks) to the time-averaged mean absolute deviation curve, $\text{MAD}(\tau)$, restricted to the short-lag scaling regime of time lags up to \SI{8}{\milli\second} (the shortest resolvable lags are \SI{0.8}{\milli\second} for the experimental LFP, set by the \SI{1250}{\hertz} sampling rate, and \SI{1}{\milli\second} for the circuit model).
With a single component, the fitted exponent is the log-log slope of the band, over which the dynamics are dominated by stochastic fluctuations rather than deterministic drift from the effective potential.
To confirm that the fitted band lies within the scaling regime, we separately locate the crossover lag as the first transition point of an unconstrained two-component fit to the full curve, verifying that the crossover falls beyond the fitted band for at least $95\%$ of curves.
We use the mean absolute deviation, rather than the mean squared displacement or detrended fluctuation analysis, for robustness against the formally unbounded variance of the increments of a L\'evy process.

For the spectral exponent $b$, we fit the MAPPLE model to the power spectral density, $\text{PSD}(f)$, normalized by its maximum (the exponents are scale invariant).
For the experimental LFP, we fit two power-law components and two Gaussian peaks (absorbing the $\theta$ and $\gamma$ oscillations) over \SIrange{1}{500}{\hertz}, holding the interior transition fixed at \SI{3}{\hertz} and the final transition at the top of the band; the spectral exponent is the slope of the component spanning \SIrange{3}{500}{\hertz}, while the component below \SI{3}{\hertz} is a nuisance parameter absorbing the low-frequency knee.
Both transitions are held fixed because a freely fitted transition is not identifiable on these smoothly curved spectra, settling near the $\gamma$ bend on some channels and near the knee on others, and because a final transition fitted inside the band would let the model roll off through its closing window rather than through its exponent.
For the circuit and bFNS models, whose spectra lack a low-frequency knee within the fitted band, we instead fit a single power-law component over \SIrange{10}{1000}{\hertz}, with one Gaussian peak absorbing the $\gamma$ oscillation of the circuit model.

For the Fano factor, we first discretize the spike times into bins of width $\tau$ spanning the recording period, and count the number of spikes occurring within each bin.
The Fano factor at timescale $\tau$ is then computed as the ratio of the variance to the mean of these spike counts across all bins, providing a measure of the variability of neural activity relative to a Poisson process (for which the Fano factor equals unity).
The variability exponent $c$ is extracted from the rising portion of the Fano factor curve, which is bounded below by a Poissonian floor at short windows.
Every Fano factor fit is a MAPPLE model with no peaks whose first power-law component is pinned to an exponent of zero to represent the Poissonian floor, with the transition width held at $0.1$ decades; each model is refined from four random restarts drawn from a fixed seed, in addition to the two refinements described above, and the lowest-objective solution is retained.
The components above the floor differ between the experimental data and the models, since the Fano factor of the circuit and mean-field neurons can saturate within the fitted window, whereas the experimental Fano factor continues to rise (\cref{subfig:empirical_fano_factor}).
For the experimental data, we fit a single free component that rises from a fitted transition to the longest window, and take $c$ as its exponent.
We prefer this single rise to a model that selects one of several rising segments, because the experimental curves rise in two stages whose relative steepness varies across areas; a selected segment would therefore measure different timescales in different areas and could not be compared along the hierarchy.
For the circuit model and the mean-field sweeps, we instead allow up to two free components above the floor, select their number by the Bayesian information criterion, and take $c$ as the exponent of the steepest rising component and that component's span as the fitted band, so that the scaling band is measured separately from the saturation that follows it.

Since the Fano factor curves of individual units and neurons are too noisy for a fit with free transitions, we fit median-aggregated curves.
For the experimental data, we compute Fano factors over $200$ logarithmically spaced windows from \SI{1}{\milli\second} to \SI{10}{\second}, take the median curve across the units of each session, area, and layer (requiring at least five units that pass the quality criteria of \nameref{subsec:experimental_data}), and fit each such curve separately; we report the median exponent across sessions, with a 95\% percentile bootstrap interval over $10^4$ resamples of sessions.
For the circuit model we fit the across-neuron median curve, and for the mean-field sweeps the across-repeat median curve of each parameter combination (\cref{subfig:variability_exponent_mean_field}); refitting the median curves of disjoint halves of the circuit neurons reproduces the circuit exponent to within $2 \times 10^{-3}$.

\subsection{Hierarchical correlation analysis}
\label{subsec:hierarchical_correlation}

To quantify hierarchical trends, we compute Kendall's $\tau$ rank correlation between each dynamical exponent and the anatomical hierarchy score of each area.
We choose $\tau$ over parametric correlation coefficients because it measures monotonic association without assuming a particular functional form, and remains robust to the heavy tails of the exponent distributions.
We report the across-session median $\tau$ at each of four cortical layers (2/3, 4, 5, and 6), with 95\% percentile bootstrap intervals over $10^4$ resamples of sessions.
A session contributes a $\tau$ at a given layer only if at least four of the six areas carry an exponent there; this requirement binds mainly for the variability exponent, whose per-area fits need at least five units, leaving between $51$ and $69$ sessions per layer against $68$ for the diffusion and spectral exponents.
Significance is assessed by a permutation test that shuffles the hierarchy scores independently within each session, using the across-session mean $\tau$ as the test statistic.
We compute $p$ over $10^4$ permutations with the add-one estimator, which counts the observed data among its own permutations and so cannot return zero, and adjust across the twelve exponent-by-layer cells by the Benjamini--Hochberg procedure, as detailed in~\textcite{Harris2026}.

\nolinenumbers
\clearpage
\bibliography{theBibliography}

\end{document}


\captionsetup{labelfont=bf,singlelinecheck=false,justification=raggedright,font={stretch=1,footnotesize}}

\title{\mytitle}

\author{Brendan Harris}
\affiliation{School of Physics, The University of Sydney, Camperdown NSW 2006, Australia}

\author{Pulin Gong}
\affiliation{School of Physics, The University of Sydney, Camperdown NSW 2006, Australia}
\affiliation{ARC Centre of Excellence for Integrative Brain Function, The University of Sydney, Camperdown NSW 2006, Australia}

\onecolumngrid
\begin{center} \huge{\textsc{Supplemental material}}
\end{center}
\maketitle
\onecolumngrid
\vspace{-2em}
\clearpage


\begin{figure*}[htbp]
	\centering
	\includegraphics[width=\textwidth]{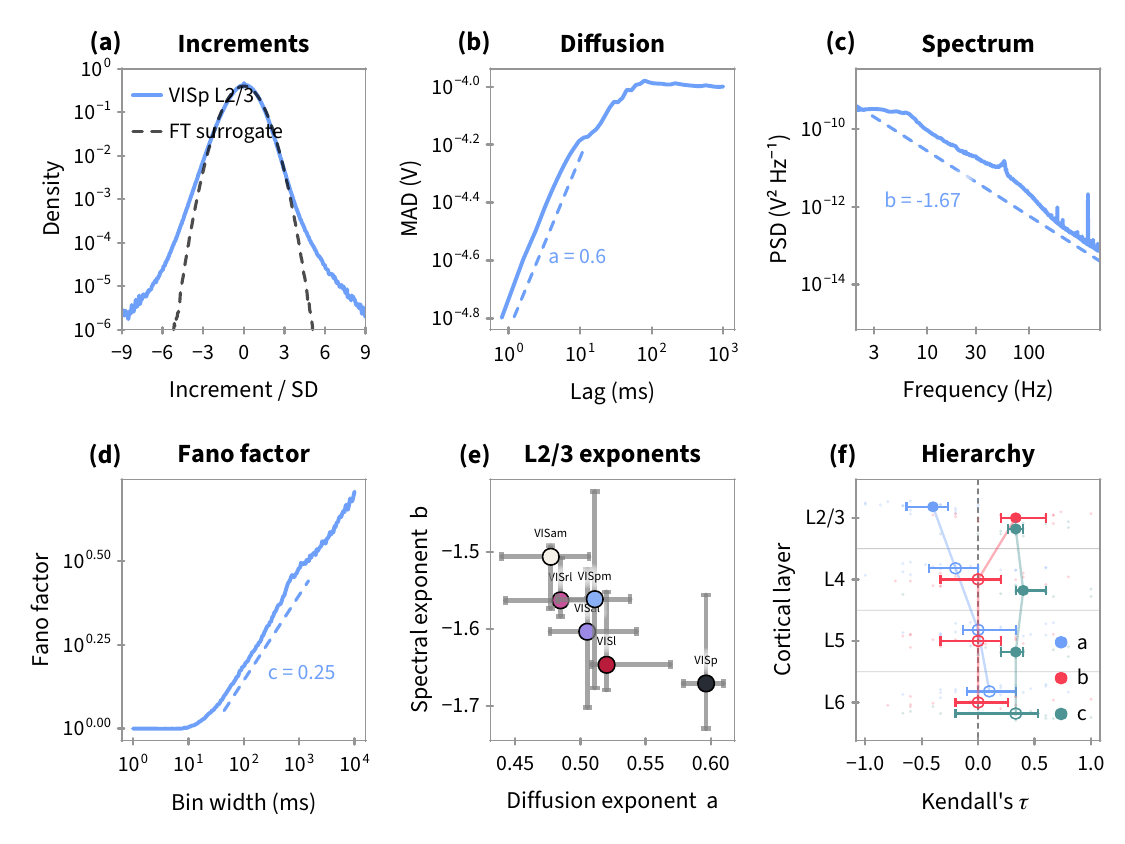}
	{
		\phantomsubcaption\label{subfig:visual_coding_increments}
		\phantomsubcaption\label{subfig:visual_coding_mad}
		\phantomsubcaption\label{subfig:visual_coding_psd}
		\phantomsubcaption\label{subfig:visual_coding_fano}
		\phantomsubcaption\label{subfig:visual_coding_exponents}
		\phantomsubcaption\label{subfig:visual_coding_hierarchy}
	}
	\caption{The adaptive fractional state in an independent Neuropixels dataset.
	All panels are computed from the Allen Institute Visual Coding dataset, using $23$ sessions of the `Functional Connectivity' stimulus set.
	The `Functional Connectivity' stimulus set includes a single $\sim$\SI{30}{\minute} spontaneous epoch, which is longer the $\sim$\SI{5}{\minute} spontaneous epochs in the `Visual Behavior' data presented in the main text but for fewer subjects.
	\newsubref{subfig:visual_coding_increments} Distribution of LFP increments (after standardizing channels) pooled across layer 2/3 channels of \texttt{VISp} (blue) against a phase-randomized (FT) surrogate of the same channels that preserves each power spectrum but enforces Gaussian increments (dashed black), as in \cref{subfig:lfp_increments}.
	The empirical distribution retains heavy tails out to $\pm 9$ standard deviations, whereas the surrogate decays as a Gaussian.
	\newsubref{subfig:visual_coding_mad} Mean absolute deviation (MAD) of the \texttt{VISp} layer 2/3 LFP, as in \cref{subfig:empirical_mad}.
	The dashed line shows the across-session median exponent over the short-lag window (up to \SI{8}{\milli\second}) over which the diffusion exponent is fitted.
	\newsubref{subfig:visual_coding_psd} Power spectral density of the same recordings, with the dashed line showing the across-session median aperiodic exponent, as in \cref{subfig:empirical_psd}.
	The spectrum is substantially shallower than $f^{-2}$ ($b = -1.67$), indicating long-range temporal dependence.
	\newsubref{subfig:visual_coding_fano} Fano factor of \texttt{VISp} layer 2/3 units against the spike-counting window, as in \cref{subfig:empirical_fano_factor}, with the dashed line showing the across-session median variability exponent (drawn over \SIrange{30}{1000}{\milli\second}).
	Spiking is Poissonian ($F \approx 1$) below $\sim$\SI{10}{\milli\second} and becomes super-Poissonian with a power-law increase in Fano factor at longer windows ($c = 0.25$).
	\newsubref{subfig:visual_coding_exponents} Diffusion exponent $a$ and spectral exponent $b$ in layer 2/3 for each of the six visual areas, colored by anatomical hierarchy score as in \cref{subfig:cortex_map}.
	Points give the across-session bootstrap median and bars the 95\% confidence interval.
	As in \cref{subfig:hierarchy_l23}, \texttt{VISp} sits at the most superdiffusive end of the axis, with higher areas falling toward normal diffusion and shallower spectra.
	\newsubref{subfig:visual_coding_hierarchy} Kendall's $\tau$ correlation between each dynamical exponent and the anatomical hierarchy score, by cortical layer, for the diffusion exponent $a$ (blue), the spectral exponent $b$ (red), and the variability exponent $c$ (teal).
	Filled markers indicate a Benjamini--Hochberg-adjusted permutation $p < 0.01$, as in \cref{subfig:hierarchy_correlation}.
	\label{fig:visual_coding}
	}
\end{figure*}

\begin{figure*}[htbp]
	\centering
	\includegraphics[width=\textwidth]{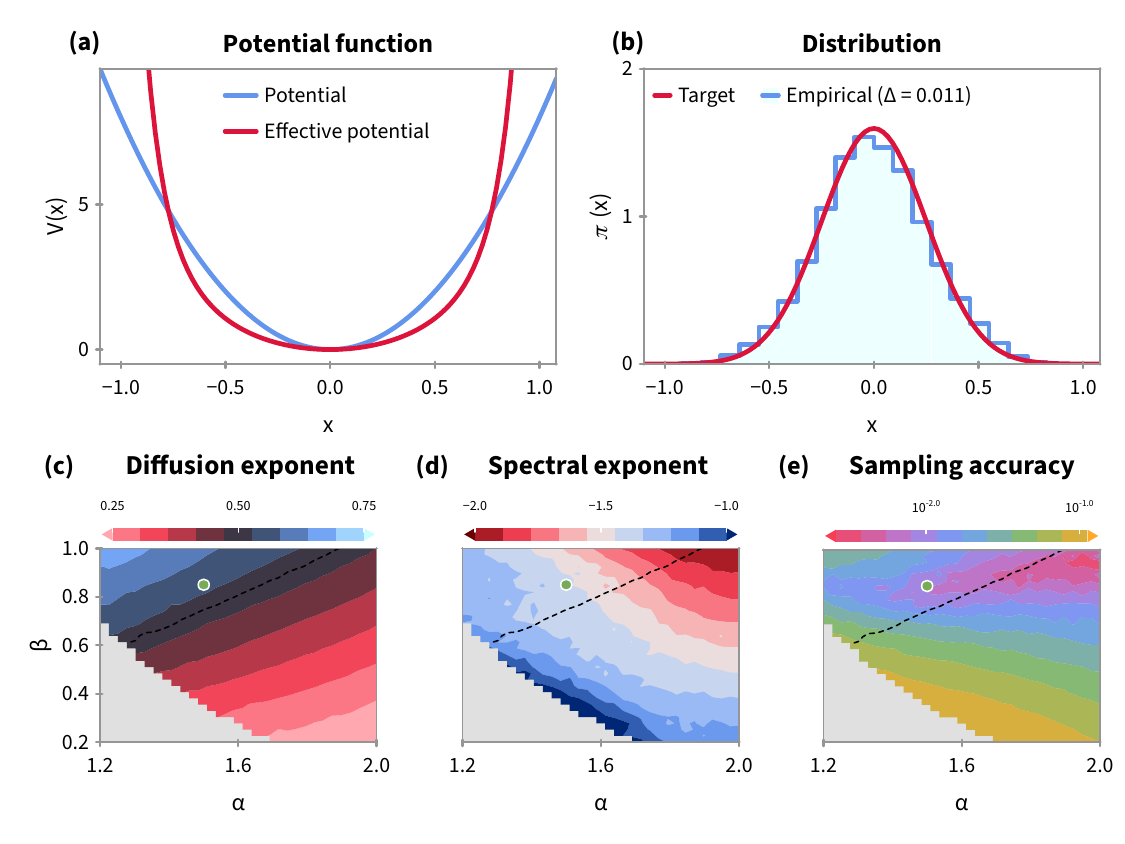}
	{
		\phantomsubcaption\label{subfig:effective_potential_unimodal}
		\phantomsubcaption\label{subfig:stationary_distribution_unimodal}
		\phantomsubcaption\label{subfig:diffusion_exponent_unimodal}
		\phantomsubcaption\label{subfig:spectral_exponent_unimodal}
		\phantomsubcaption\label{subfig:sampling_accuracy_unimodal}
	}
	\caption{Dynamical exponents and sampling accuracy for bFNS on a unimodal potential.
	\newsubref{subfig:effective_potential_unimodal} The potential function $V(x) = -\log \pi(x)$ (blue) for a unimodal target distribution $\pi(x)$, along with the effective potential for spatial fractional order $\alpha = 1.5$ (red).
	The fractional spatial derivative flattens interior regions of the potential and increases the steepness of the edge regions.
	\newsubref{subfig:stationary_distribution_unimodal} Target distribution $\pi(x)$ (red line) and the empirical distribution of samples generated by the bi-fractional model (blue histogram) with parameters $\alpha = 1.5$, $\beta = 0.85$, $\gamma=0.03$, and $\eta = 0.01$. The Wasserstein distance $\Delta$ between the target and empirical distributions is annotated.
	\newsubref{subfig:diffusion_exponent_unimodal} Heatmap of diffusion exponents $a$ across spatial fractional order $\alpha$ and temporal fractional order $\beta$.
	Gray indicates invalid parameter combinations where the self-similarity exponent lies outside $(0, 1)$.
	\newsubref{subfig:spectral_exponent_unimodal} Heatmap of spectral exponents $b$ across $\alpha$ and $\beta$. The black dashed line indicates $a = 0.5$ from \subref{subfig:diffusion_exponent_unimodal}.
	\newsubref{subfig:sampling_accuracy_unimodal} Heatmap of sampling accuracy (Wasserstein distance between target and empirical distributions) across $\alpha$ and $\beta$.
	\label{fig:theoretical_model_unimodal}
	}
\end{figure*}


\begin{figure*}[htbp]
	\centering
	\includegraphics[width=\textwidth]{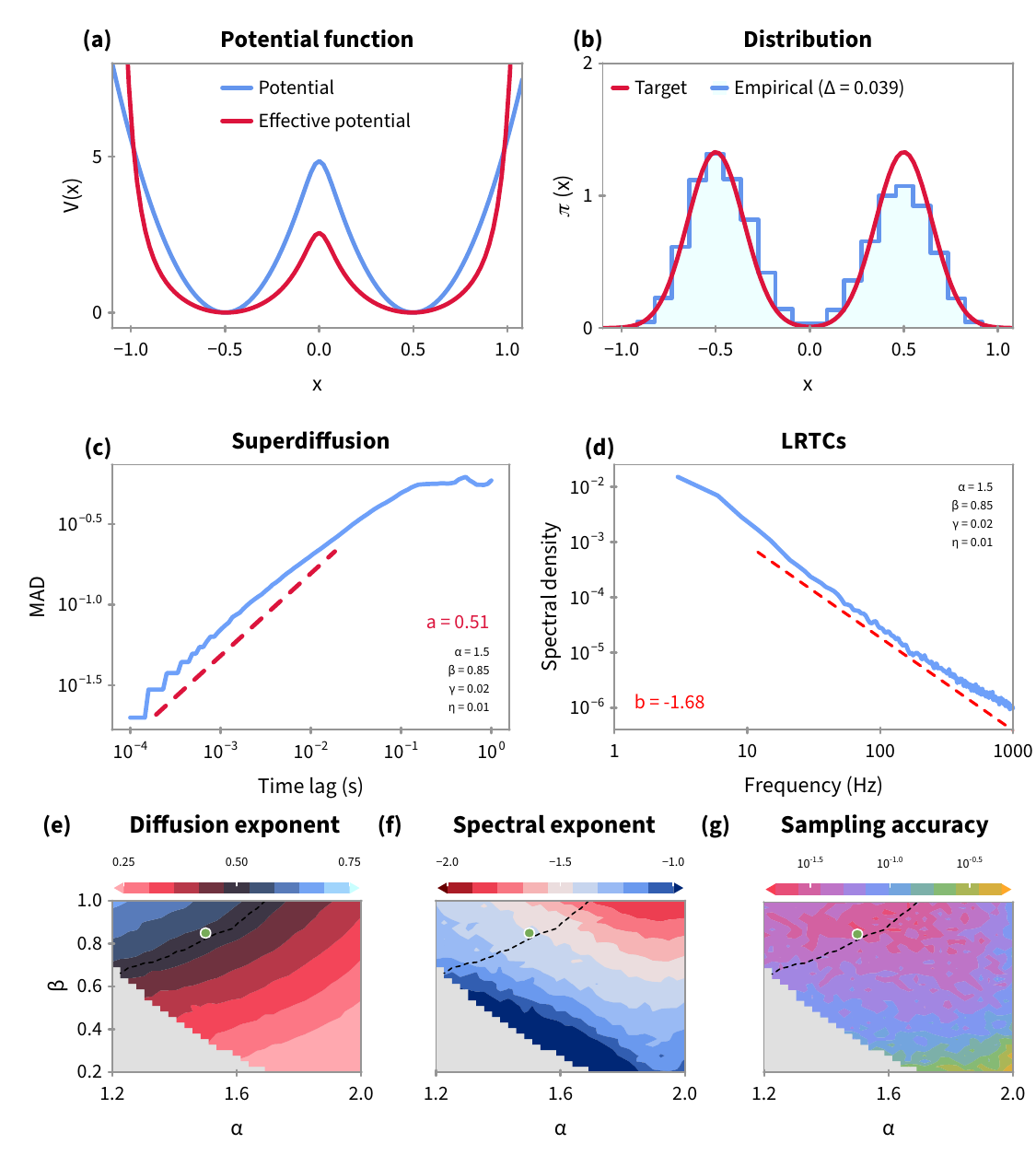}
	{
		\phantomsubcaption\label{subfig:effective_potential_bimodal}
		\phantomsubcaption\label{subfig:stationary_distribution_bimodal}
		\phantomsubcaption\label{subfig:diffusion_exponent_bimodal}
		\phantomsubcaption\label{subfig:spectral_exponent_bimodal}
		\phantomsubcaption\label{subfig:diffusion_exponent_sweep_bimodal}
		\phantomsubcaption\label{subfig:spectral_exponent_sweep_bimodal}
		\phantomsubcaption\label{subfig:sampling_accuracy_bimodal}
	}
	\vspace{-2em}
	\caption{bFNS dynamics on a bimodal potential.
	\newsubref{subfig:effective_potential_bimodal} The double-well potential function $V(x)$ (blue) for a bimodal target distribution, along with the effective potential for spatial fractional order $\alpha = 1.5$ (red).
	The fractional spatial derivative reduces the barrier height between modes and increases the steepness of the extreme edges.
	\newsubref{subfig:stationary_distribution_bimodal} Target distribution $\pi(x)$ (red line) for the bimodal potential in \subref{subfig:effective_potential_bimodal}, along with the empirical distribution of samples generated by the bi-fractional model (histogram) with parameters $\alpha = 1.5$, $\beta = 0.85$, $\gamma = 0.02$, and $\eta = 0.01$. The Wasserstein distance $\Delta$ between the target and empirical distributions is annotated.
	\newsubref{subfig:diffusion_exponent_bimodal} Mean absolute deviation (MAD) of the bi-fractional model with parameters in \subref{subfig:stationary_distribution_bimodal}. The diffusion exponent $a$ is annotated.
	\newsubref{subfig:spectral_exponent_bimodal} Power spectral density of the process in \subref{subfig:stationary_distribution_bimodal}. The spectral exponent $b$ is annotated.
	\\[0.5em](continued)
	}
	\end{figure*}
	\begin{figure*}
	\ContinuedFloat
	\centering
	\caption{
	\newsubref{subfig:diffusion_exponent_sweep_bimodal} Heatmap of diffusion exponents $a$ for the bi-fractional model with a bimodal potential across spatial fractional order $\alpha$ and temporal fractional order $\beta$. The gray area corresponds to invalid parameter combinations where the self-similarity exponent lies outside $(0, 1)$. The green dot indicates the parameter combination used in \cref{subfig:effective_potential_bimodal,subfig:stationary_distribution_bimodal,subfig:diffusion_exponent_bimodal,subfig:spectral_exponent_bimodal}.
	\newsubref{subfig:spectral_exponent_sweep_bimodal} Heatmap of spectral exponents $b$ across $\alpha$ and $\beta$. The black dashed line indicates the boundary where $a = 0.5$.
	\newsubref{subfig:sampling_accuracy_bimodal} Heatmap of sampling accuracy across $\alpha$ and $\beta$.
	\label{fig:theoretical_model_bimodal}
	}
\end{figure*}

\begin{figure*}[htbp]
	\centering
	\includegraphics[width=\textwidth]{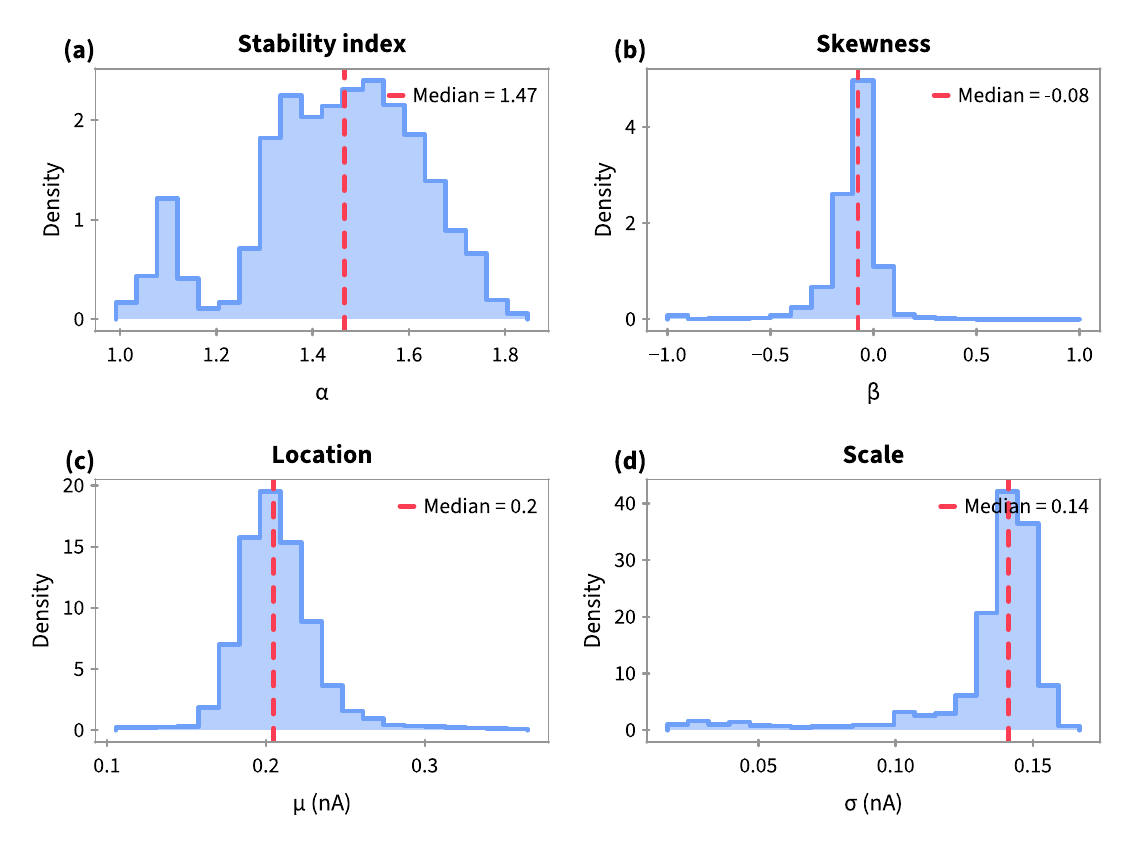}
	{
		\phantomsubcaption\label{subfig:input_alpha}
		\phantomsubcaption\label{subfig:input_beta}
		\phantomsubcaption\label{subfig:input_mu}
		\phantomsubcaption\label{subfig:input_sigma}
		}
	\caption{Parameters of a stable distribution fit to the synaptic input currents for each neuron in the circuit model.
	\newsubref{subfig:input_alpha} The tail (stability) index $\alpha$, which determines the limiting power-law exponent of the distribution tails. A value of $\alpha = 2$ corresponds to a Gaussian distribution.
	\newsubref{subfig:input_beta} The skewness parameter $\beta$, which determines the asymmetry of the distribution. A value of $\beta = 0$ corresponds to a symmetric distribution.
	\newsubref{subfig:input_mu} The location parameter $\mu$, which determines the center location of the distribution.
	\newsubref{subfig:input_sigma} The scale parameter $\sigma$, which determines the spread of the distribution.
	}
	\label{sup:input_distribution_parameters}
\end{figure*}

\clearpage
\begin{video}[htbp]
	\caption{Wandering of the localized activity pattern in the circuit model.
	The total synaptic input current is shown across the $\SI{0.5}{\milli\meter}\times\SI{0.5}{\milli\meter}$ network, with color giving the input current in \unit{\nano\ampere}.
	A black dot marks the center of mass of network activity, traced for \SI{40}{\milli\second} by a colored line, which shows intermittent superdiffusive jumps.
	}
	\label{video:input_field}
\end{video}
\clearpage
